\documentclass[orivec]{llncs}
\usepackage[firstpage]{draftwatermark}

\usepackage[utf8]{inputenc}

\usepackage{underscore}

\usepackage{cite}
\usepackage{paralist}
\usepackage{mathtools}
\usepackage{xspace}
\usepackage{amssymb}
\usepackage{amsmath}
\usepackage{amsthm}
\usepackage{thmtools}
\usepackage{wrapfig}
\usepackage{bm}
\usepackage{hyperref}
\usepackage[dvipsnames]{xcolor}

\newcommand{\setn}[1]{[#1]}
\newcommand{\naturals}{\mathbb{N}}
\newcommand{\setnocond}[1]{\{#1\}}
\newcommand{\setcond}[2]{\{\, #1 \mid #2 \,\}}

\newcommand{\orderPosition}{\mathsf{ord}}

\newcommand{\wordletter}[2]{{#1}[#2]}

\newcommand{\vertex}[2]{\langle {#1}, {#2} \rangle}
\newcommand{\edges}[1][E]{#1}
\newcommand{\vertices}[1][V]{#1}

\newcommand{\graph}[2]{\langle {#1}, {#2} \rangle}
\newcommand{\branch}{\gamma}

\newcommand{\alphabet}{\Sigma}

\newcommand{\finwords}{\alphabet^*}
\newcommand{\infwords}{\alphabet^{\omega}}

\newcommand{\langsymb}[0]{\mathsf{L}}
\newcommand{\lang}[1]{\langsymb(#1)}

\newcommand{\delaSymbol}{\mathsf{E}}
\newcommand{\dela}[1][A]{\aut[#1]^{\delaSymbol}}
\newcommand{\delaStates}{\states^{\delaSymbol}}
\newcommand{\delaTrans}{\trans^{\delaSymbol}}
\newcommand{\delaInitial}{\iota^{\delaSymbol}}
\newcommand{\delaColor}{\Gamma^{\delaSymbol}}
\newcommand{\delaParity}{\parity^{\delaSymbol}}
\newcommand{\delaAcc}{\acc^{\delaSymbol}}

\newcommand{\basenumber}{\mathsf{b}}

\newcommand{\colorsymbol}[1]{\Gamma_{#1}}
\newcommand{\accinf}[1]{\mathsf{Inf}(#1)}
\newcommand{\accfin}[1]{\mathsf{Fin}(#1)}
\newcommand{\acctrue}{\mathsf{tt}}
\newcommand{\accfalse}{\mathsf{ff}}
\newcommand{\accformula}{\alpha}

\newcommand{\reachweak}{\mathsf{W}}
\newcommand{\reachweakacc}{\mathsf{WA}}
\newcommand{\reachdet}{\mathsf{D}}
\newcommand{\reachnondet}{\mathsf{N}}

\newcommand{\snode}{\mathsf{node}}
\newcommand{\states}{Q}

\newcommand{\dstates}{\states_{D}}
\newcommand{\trans}{\delta}
\newcommand{\cotrans}[1][e]{\delta^{e}}
\newcommand{\ntrans}{\delta_{N}}

\newcommand{\dtrans}{\delta_{D}}

\newcommand{\init}{\iota}
\newcommand{\acc}{\mathsf{Acc}}
\newcommand{\run}{\rho}

\newcommand{\accset}{F}

\newcommand{\rabinSymbol}{\mathsf{R}}

\newcommand{\dra}[1][A]{\aut[#1]^{\rabinSymbol}}

\newcommand{\parity}{\mathsf{p}}
\newcommand{\paritySymbol}{\mathsf{P}}

\newcommand{\parityTranAccValues}{\mathsf{G}}
\newcommand{\parityTranRejValues}{\mathsf{B}}

\newcommand{\dpa}[1][A]{\aut[#1]^{\paritySymbol}}

\newcommand{\mystackrelsingle}[2]{%
  \mathrel{\vbox{\offinterlineskip\ialign{%
    \hfil##\hfil\cr%
    $\scriptstyle#1$\cr%
    \noalign{\kern.3ex}%
    $#2$\cr%
}}}}
\newcommand{\transition}[3]{#1 \mystackrelsingle{#2}{\longrightarrow} #3}

\newcommand{\stateOrder}{\preccurlyeq}

\newcommand{\dagAW}[2]{\mathcal{G}_{#1, #2}}

\newcommand{\aut}[1][A]{\mathcal{#1}}

\newcommand{\bigO}{\mathsf{O}}
\newcommand{\bigM}{\mathsf{\Theta}}

\newcommand{\labellingtree}{\mathfrak{t}}

\newcommand{\labelling}{\mathfrak{g}}

\newcommand{\labellingUndef}{\infty}

\newcommand{\values}[1]{\beta(#1)}

\newcommand{\listsnac}[1][\reachnondet]{\mathcal{L}_{#1}}
\newcommand{\listnac}[1]{[#1]}
\newcommand{\listconcat}[2]{#1 {\smallfrown} #2}

\newcommand{\buchi}{B\"uchi\xspace}

\newcommand{\size}[1]{|#1|}
\newcommand{\inftimes}[1]{\mathit{inf}({#1})}

\newcommand{\automatabenchmarks}{\textsc{automata-benchmarks}\xspace}
\newcommand{\benchexec}{\textsc{BenchExec}\xspace}
\newcommand{\spot}{\textsc{Spot}\xspace}
\newcommand{\owl}{\textsc{Owl}\xspace}

\newcommand{\cola}{\textsc{COLA}\xspace}
\newcommand{\autfilt}{\texttt{autfilt}\xspace}

\newcommand{\weaksucc}{\mathsf{weakSucc}}
\newcommand{\detsucc}{\mathsf{detSucc}}
\newcommand{\nondetsucc}{\mathsf{nondetSucc}}

\allowdisplaybreaks

\SetWatermarkAngle{0}
\SetWatermarkText{
 \vspace{-2.5cm}
 \raisebox{12.5cm}{%
  \hspace{-0.1cm}%
  \href{https://doi.org/10.5281/zenodo.6558928}{\includegraphics{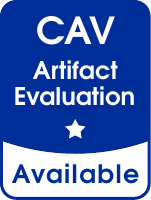}}%
  \hspace{8.6cm}%
  \includegraphics{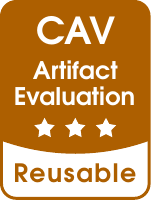}%
}}

\title{Divide-and-Conquer Determinization of B\"uchi Automata Based on SCC Decomposition}
\author{
Yong Li\inst{1}
\orcidID{0000-0002-7301-9234}
\and
Andrea Turrini\inst{1,2}
\orcidID{0000-0003-4343-9323}
\and
Weizhi Feng\inst{1,3}
\orcidID{0000-0003-0710-223X}
\and
\\
Moshe Y. Vardi\inst{4}
\orcidID{0000-0002-0661-5773}
\and
Lijun Zhang\inst{1,2}
\orcidID{0000-0002-3692-2088}
}

\institute{
State Key Laboratory of Computer Science,
Institute of Software, Chinese Academy of Sciences, China
\and
Institute of Intelligent Software Guangzhou, China
\and
University of Chinese Academy of Sciences, China
\and
Rice University, USA
}

\begin{document}
\pagestyle{plain}

\maketitle

\begin{abstract}
The determinization of a nondeterministic \buchi automaton (NBA) is a fundamental construction of automata theory, with applications to probabilistic verification and reactive synthesis.
The standard determinization constructions, such as the ones based on the Safra-Piterman's approach, work on the whole NBA.
In this work we propose a divide-and-conquer determinization approach.
To this end, we first classify the strongly connected components (SCCs) of the given NBA as inherently weak, deterministic accepting, and nondeterministic accepting.
We then present how to determinize each type of SCC \emph{independently} from the others;
this results in an easier handling of the determinization algorithm that takes advantage of the structure of that SCC.
Once all SCCs have been determinized, we show how to compose them so to obtain the final equivalent deterministic Emerson-Lei automaton, which can be converted into a deterministic Rabin automaton without blow-up of states and transitions.
We implement our algorithm in a our tool \cola and empirically evaluate \cola with the state-of-the-art tools \spot and \owl on a large set of benchmarks from the literature.
The experimental results show that our prototype \cola outperforms \spot and \owl regarding the number of states and transitions.

\end{abstract}

\section{Introduction}
\label{sec:intro}
Nondeterministic \buchi automata (NBAs)~\cite{buchi90decision} are finite automata accepting infinite words;
they are a simple and popular formalism used in model checking to represent reactive and non-terminating systems and their specifications, characterized by $\omega$-regular languages~\cite{baier2008principles}.
Due to their nondeterminism, however, there are situations in which NBAs are not suitable, so deterministic automata are required, as it happens in probabilistic verification~\cite{baier2008principles} and reactive synthesis from logical specifications~\cite{DBLP:conf/popl/PnueliR89}.
Consequently, translating NBAs into equivalent deterministic $\omega$-automata (that is, deterministic automata accepting the same $\omega$-regular language) is a necessary operation for solving these problems.
While there exists a direct translation from linear temporal logic (LTL) to deterministic $\omega$-automata~\cite{Esparza20JACM}, not all problems of interests can be formalized by LTL formulas, since LTL cannot express the full class of $\omega$-regular properties~\cite{DBLP:journals/iandc/VardiW94}.
For instance, we have to use Linear Dynamic Logic (LDL)~\cite{DBLP:conf/gandalf/Vardi11,DBLP:conf/ijcai/GiacomoV13} instead of LTL to express the $\omega$-regular property ``the train will arrive in every odd minute".
To the best of our knowledge, we still need to go through the determinization of NBAs for LDL to obtain deterministic $\omega$-automata.
Therefore, NBA determinization is very important in verifying the whole class of $\omega$-regular properties.

The determinization of NBAs is a fundamental problem in automata theory that has been actively studied for decades.
For the determinization of nondeterministic automata accepting finite words, it suffices to use a subset construction~\cite{Hopcroft2006IAT}.
Determinization constructions for NBAs are, however, much more involved since the simple subset construction is not sufficient~\cite{safra1988complexity}.
Safra~\cite{safra1988complexity} gave the first determinization construction for NBAs with the optimal complexity $2^{\bigO(n \log n)}$, here $n$ is the number of states of the input NBA;
Michel~\cite{Michel/88/lowerComplementation} then gave a lower bound $n!$ for determinizing NBAs.
Safra's construction has been further optimized by Piterman~\cite{DBLP:journals/lmcs/Piterman07} to $\bigO((n!)^{2})$~\cite{Schewe09}, resulting in the widely known Safra-Piterman's construction.
The Safra-Piterman's construction is rather challenging, while still being the most practical way for \buchi complementation~\cite{TsaiFVT14}.
Research on determinization since then either aims at developing alternative Safraless constructions~\cite{kahler2008complementation,DBLP:journals/iandc/FogartyKVW15,DBLP:conf/icalp/LodingP19} or further tightening the upper and lower bounds of the NBA determinization~\cite{DBLP:conf/icalp/ColcombetZ09,Yan/08/lowerComplexity,DBLP:conf/fossacs/Schewe09,DBLP:journals/ipl/LiuW09}.

In this paper, we focus on the \emph{practical} aspects of \buchi determinization.
All works on determinization mentioned above focus on translating NBAs to either deterministic Rabin or deterministic parity automata.
According to~\cite{DBLP:conf/stoc/SafraV89}, the more relaxed an acceptance condition is, the more succinct a finite automaton can be, regarding the number of states.
In view of this, we consider the translation of NBAs to deterministic \emph{Emerson-Lei} automata (DELAs)~\cite{DBLP:journals/scp/EmersonL87, DBLP:conf/stoc/SafraV89} whose acceptance condition is an arbitrary Boolean combination of sets of transitions to be seen finitely or infinitely often, the most generic acceptance condition for a deterministic automaton.
We consider here transition-based automata rather than the usual state-based automata since the former can be more succinct~\cite{duret2016spot}.

The \buchi determinization algorithms available in literature operate on the \emph{whole} NBA structure at once, which does not scale well in practice due to the complex structure and the big size of the input NBA.
In this work we apply a \emph{divide-and-conquer} methodology to \buchi determinization.
We propose a determinization algorithm for NBAs to DELAs based on their strongly connected components (SCCs) decomposition.
We first classify the SCCs of the given NBA into three types:
\emph{inherently weak}, in which either all cycles do not visit accepting transitions or all must visit accepting transitions;
\emph{deterministic accepting} and \emph{nondeterministic accepting}, which contain an accepting transition and are deterministic or nondeterministic, respectively.
We show how to divide the whole \buchi determinization problem into the determinization for each type of SCCs \emph{independently}, in which the determinization for an SCC takes advantage of the structure of that SCC.
Then we show how to compose the results of the local determinization for each type of SCCs, leading to the final equivalent DELA.
An extensive experimental evaluation confirms that the divide-and-conquer approach pays off also for the determinization of the whole NBA.

\paragraph*{\bf Contributions.}
First, we propose a \emph{divide-and-conquer} determinization algorithm for NBAs, which takes advantage of the structure of different types of SCCs and determinizes SCCs independently.
Our construction builds an equivalent DELA that can be converted into a deterministic Rabin automaton without blowing up states and transitions (cf. Theorem~\ref{thm:dela-to-dra}).
To the best of our knowledge, we propose the \emph{first} determinization algorithm that constructs a DELA from an NBA.
Second, we show that there exists a family of NBAs for which our algorithm gives a DELA of size $2^{n+2}$ while classical works construct a DPA of size at least $n!$ (cf. Theorem~\ref{thm:better-complexity-nbas}).
Third, we implement our algorithm in our tool \cola and evaluate it with the state-of-the-art tools \spot~\cite{duret2016spot} and \owl~\cite{DBLP:conf/atva/KretinskyMS18} on a large set of benchmarks from the literature.
The experiments show that \cola outperforms \spot and \owl regarding the number of states and transitions.
Finally, we remark that the determinization complexity for some classes of NBAs can be exponentially better than the known ones (cf. Corollary~\ref{coro:worst-case-subclasses-determinization}).

We defer all proofs to Appendix~\ref{app:proofs}.
\section{Preliminaries}
\label{sec:preliminaries}

Let $\alphabet$ be a given alphabet, i.e., a finite set of letters.
A transition-based Emerson-Lei automaton
can be seen as a generalization of other types of $\omega$-automata, like \buchi, Rabin or parity.
Formally, it is defined in the HOA format~\cite{Babiak15} as follows:
\begin{definition}
\label{def:automaton}
    A \emph{nondeterministic Emerson-Lei automaton} (NELA) is a tuple $\aut = (\states, \init, \trans, \colorsymbol{k}, \parity, \acc)$, where
    $\states$ is a finite set of \emph{states};
    $\init \in \states$ is the \emph{initial state};
    $\trans \subseteq \states \times \alphabet \times \states$ is a \emph{transition relation};
    $\colorsymbol{k} = \setnocond{0, 1, \cdots, k}$, where $k \in \naturals$, is a set of \emph{colors};
    $\parity \colon \trans \to 2^{\colorsymbol{k}}$ is a \emph{coloring function} for transitions;
    and
    $\acc$ is an  \emph{acceptance formula} over $\colorsymbol{k}$ given by the following grammar, where $x \in \colorsymbol{k}$:
    \[
        \accformula := \acctrue \mid \accfalse \mid \accfin{x} \mid \accinf{x} \mid \accformula \lor \accformula \mid \accformula \land \accformula.
    \]
\end{definition}
We remark that the colors in $\colorsymbol{k}$ are not required to be all used in $\acc$.
We call a NELA a \emph{deterministic} Emerson-Lei automaton (DELA) if for each $q \in \states$ and $a \in \alphabet$, there is at most one $q' \in \states$ such that $(q,a, q') \in \trans$.

In the remainder of the paper, we consider $\trans$ also as a function $\trans \colon \states \times \alphabet \to 2^{\states}$ such that $q' \in \trans(q, a)$ whenever $(q, a, q') \in \trans$;
we also write $\transition{q}{a}{q'}$ for $(q, a, q') \in \trans$ and we extend it to words $u = u_{0} u_{1} \cdots u_{n} \in \finwords$ in the natural way, i.e., $\transition{q}{u}{q'} = \transition{q}{\wordletter{u}{0}}{q_{1}} \transition{}{\wordletter{u}{1}}{} \cdots \transition{}{\wordletter{u}{n}}{q'}$, where $\wordletter{\sigma}{i}$ denotes the element $s_{i}$ of the sequence of elements $\sigma = s_{0} s_{1} s_{2} \cdots$ at position $i$.
We assume without loss of generality that each automaton is \emph{complete}, i.e., for each state $q \in \states$ and letter $a \in \alphabet$, we have $\trans(q, a) \neq \emptyset$.
If it is not complete, we make it complete by adding a fresh state $q_{\bot} \notin \states$ and redirecting all missing transitions to it.

A \emph{run} of $\aut$ over an $\omega$-word $w \in \infwords$ is an infinite sequence of states $\run$ such that $\wordletter{\run}{0} = \init$, and for each $i \in \naturals$, $(\wordletter{\run}{i}, \wordletter{w}{i}, \wordletter{\run}{i + 1}) \in \trans$.

The \emph{language} $\lang{\aut}$ of $\aut$ is the set of words accepted by $\aut$, i.e., the set of words $w \in \infwords$ such that there exists a run $\run$ of $\aut$ over $w$ such that $\parity(\inftimes{\run}) \models \acc$, where $\inftimes{\run} = \setcond{(q, a, q') \in \trans}{\forall i \in \naturals. \exists j > i. (\wordletter{\run}{j}, \wordletter{w}{j}, \wordletter{\run}{j + 1}) = (q, a, q')}$ and the satisfaction relation $\models$ is defined recursively as follows:
given $M \subseteq \colorsymbol{k}$,
\begin{align*}
M & \models \acctrue, & M &\models \accfin{x} \text{ iff $x \notin M$,}  & M & \models \accformula_{1} \lor \accformula_{2} \text{ iff $M \models  \accformula_{1}$ or $M \models \accformula_{2}$,} \\
M & \not\models \accfalse, & M &\models \accinf{x} \text{ iff $x \in M$,} & M & \models \accformula_{1} \land \accformula_{2} \text{ iff $M \models  \accformula_{1}$ and $M \models \accformula_{2}$.}
\end{align*}
Intuitively, a run $\run$ over $w$ is accepting if the set of colors (induced by $\parity$) that occur infinitely often in $\run$ satisfies the acceptance formula $\acc$.
Here $\accfin{x}$ specifies that the color $x$ only appears for finitely many times while $\accinf{x}$ requires the color $x$ to be seen infinitely often.

The more common types of $\omega$-automata, such as \buchi, parity and Rabin can be treated as Emerson-Lei automata with the following acceptance formulas.

\begin{definition}%[Types of $\omega$-automata by acceptance]
\label{def:typesOfAutomataByAcceptance}
    A NELA $\aut = (\states, \init, \trans, \colorsymbol{k}, \parity, \acc)$ is said to be
    \begin{itemize}
    \item
        a \emph{\buchi automaton} (BA) if $k = 0$ and $\acc = \accinf{0}$.
        Transition with color $0$ are usually called  \emph{accepting} transitions.
        Thus, a run $\run$ is accepting if $\parity(\inftimes{\run}) \cap \setnocond{0} \neq \emptyset$, i.e., $\run$ takes accepting transitions infinitely often;
    \item
        a \emph{parity automaton} (PA) if $k$ is even and
        $\acc = \bigvee_{c = 0}^{k/2} (\bigwedge_{i = 1}^{c} \accfin{2i -1} \land \accinf{2c})$.
        A run $\run$ is accepting if the minimum color in $\parity(\inftimes{\run})$ is even;
    \item
        a \emph{Rabin automaton} (RA) if $k$ is an odd number and $\acc = (\accfin{0} \land \accinf{1}) \lor \cdots \lor (\accfin{k-1} \land \accinf{k})$.
        Intuitively, a run $\run$ is accepting if there exists an odd integer $0 < j \leq k$ such that $j - 1 \notin \parity(\inftimes{\run})$ and $j \in \parity(\inftimes{\run})$.
    \end{itemize}
\end{definition}

When the NELA $\aut = (\states, \init, \trans, \colorsymbol{k}, \parity, \acc)$ is a nondeterministic BA (NBA), we just write $\aut$ as $(\states, \init, \trans, \accset)$ where $\accset$ is the set of accepting transitions.
We call a set $C \subseteq \states$ a \emph{strongly connected component} (SCC) of $\aut$ if for every pair of states $q, q' \in C$, we have that $\transition{q}{u}{q'}$ for some $u \in \finwords$ and $\transition{q'}{v}{q}$ for some $v \in \finwords$, i.e., $q$ and $q'$ can be reached by each other;
by default, each state $q \in \states$ reaches itself.
$C$ is a \emph{maximal} SCC if it is \emph{not} a proper subset of another SCC.
\emph{All} SCCs considered in the work are maximal.
We call an SCC $C$ \emph{accepting} if there is a transition $(q, a, q') \in (C \times \alphabet \times C) \cap \accset$ and \emph{nonaccepting} otherwise.
We say that an SCC $C'$ is \emph{reachable} from an SCC $C$ if there exist $q \in C$ and $q'\in C'$ such that $\transition{q}{u}{q'}$ for some $u \in \finwords$.
An SCC $C$ is \emph{inherently weak} if either every cycle going through the $C$-states visits at least one accepting transition or none of the cycles visits an accepting transition.
We say that an SCC $C$ is \emph{deterministic} if for every state $q \in C$ and $a \in \alphabet$, we have $\size{\trans(q, a) \cap C} \leq 1$.
Note that a state $q$ in a deterministic SCC $C$ can have multiple successors for a letter $a$, but at most one successor remains in $C$.

\begin{figure}[t]
    \centering
    \resizebox{\linewidth}{!}{
    \includegraphics{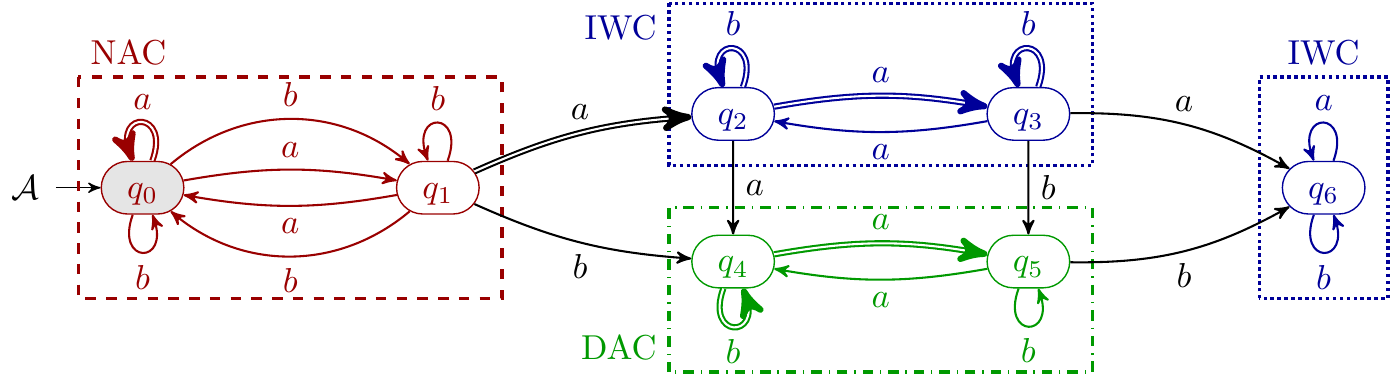}
    }
    \caption{An example of NBA.}
    \label{fig:differentSCCs}
\end{figure}

    Figure~\ref{fig:differentSCCs} shows an example of NBA we will use for our examples in the remainder of the paper;
    we depict the accepting transitions with a double arrow.
    Clearly, inside each SCC, depicted as a box, each state can be reached by any other state, and the SCCs are maximal.
    The SCC $\setnocond{q_{2}, q_{3}}$ is inherently weak and accepting, since every cycle takes an accepting transition;
    the SCC $\setnocond{q_{6}}$ is also inherently weak, but nonaccepting, since every cycle never takes an accepting transition.
    The remaining two SCCs, i.e., $\setnocond{q_{0}, q_{1}}$ and $\setnocond{q_{4}, q_{5}}$, are not inherently weak, since some cycle takes accepting transitions (like the cycle $\transition{q_{0}}{a}{q_{0}}$) while others do not (like the cycle $\transition{q_{0}}{b}{q_{0}}$).
    Both SCCs contain an accepting transition, so they are accepting;
    the SCC $\setnocond{q_{0}, q_{1}}$ is clearly nondeterministic, while the SCC $\setnocond{q_{4},q_{5}}$ is deterministic.
    Note that from $q_{5}$ we have two transitions labelled by $b$, but only the transition $\transition{q_{5}}{b}{q_{4}}$ remains inside the SCC, while the other transition $\transition{q_{5}}{b}{q_{6}}$ leaves the SCC, so the SCC is still deterministic.

The following proposition is well known and is often used in prior works.
\begin{proposition}%[Known fact]%[Forklore]
\label{prop:run-stable-scc}
    Let $\aut$ be an NBA and $w \in \infwords$.
    A run of $\aut$ over $w$ will eventually stay in an SCC.
    Moreover, if $w \in \lang{\aut}$, every accepting run of $\aut$ over $w$ will eventually stay in an accepting SCC.
\end{proposition}

Proposition~\ref{prop:run-stable-scc} is the key ingredient of our algorithm:
it allows us to determinize the SCCs independently as $\lang{\aut}$ is the union of the words whose runs stay in each accepting SCCs.
In the remainder of the paper, we first present a translation from an NBA $\aut$ to a DELA $\dela$ based on the SCC decomposition of $\aut$.
The obtained DELA $\dela$ in fact can be converted to a \emph{deterministic} Rabin automaton (DRA) $\dra$ without blowing up states and transitions, i.e., we can just convert the coloring function and the acceptance formula of $\dela$ to DRAs.

\section{Determinization Algorithms of SCCs}
\label{sec:determinizationSCCs}

Determinizing each SCC of $\aut$ independently is not straightforward since it may be reached from the initial state only after reading a nonempty finite word;
moreover, there can be words of different length leading to the SCC, entering through different states.
To keep track of the different arrivals in an SCC at different times, we make use of run DAGs~\cite{kupferman2001weak}, that are a means to organize the runs of $\aut$ over a word $w$.
In this section, we first recall the concept of run DAGs and then describe how to determinize SCCs with their help.

\begin{definition}%[Run DAG]
\label{def:runDAG}
    Let $\aut = (\states, \init, \trans, \accset)$ be an NBA and $w \in \infwords$ be a word.
    The run DAG $\dagAW{\aut}{w} = \graph{\vertices}{\edges}$ of $\aut$ over $w$ is defined as follows:
        the set of vertices $\vertices \subseteq \states \times \naturals$ is defined as $\vertices = \bigcup_{l \geq 0} (\vertices_{l} \times \setnocond{l})$ where $\vertices_{0} = \setnocond{\init}$ and $\vertices_{l + 1} = \trans(\vertices_{l}, \wordletter{w}{l})$ for every $l \in \naturals$;
        there is an edge $(\vertex{q}{l}, \vertex{q'}{l'}) \in \edges$ if $l' = l + 1$ and $q' \in \trans(q, \wordletter{w}{l})$.
\end{definition}

Intuitively, a state $q$ at a level $\ell$ may occur in several runs and only one vertex is needed to represent it, i.e., the vertex $\vertex{q}{\ell}$ who is said to be on level $\ell$.
Note that by definition, there are at most $\size{\states}$ vertices on each level.
An edge $(\vertex{q}{l}, \vertex{q'}{l + 1})$ is an \emph{$\accset$-edge} if $(q, \wordletter{w}{l}, q') \in \accset$.
An infinite sequence of vertices $\branch = \vertex{q_{0}}{0} \vertex{q_{1}}{1} \cdots$ is called an \emph{$\omega$-branch} of $\dagAW{\aut}{w}$ if $q_{0} = \init$ and for each $\ell \in \naturals$, we have $(\vertex{q_{\ell}}{\ell}, \vertex{q_{\ell + 1}}{\ell+1}) \in \edges$.
We can observe that there is a bijection between the set of runs of $\aut$ on $w$ and the set of $\omega$-branches in $\dagAW{\aut}{w}$.
In fact, to a run $\run = q_{0} q_{1} \cdots$ of $\aut$ over $w$ corresponds the $\omega$-branch $\hat{\run} = \vertex{q_{0}}{0} \vertex{q_{1}}{1} \cdots$ and, symmetrically, to an $\omega$-branch $\branch = \vertex{q_{0}}{0} \vertex{q_{1}}{1} \cdots$ corresponds the run $\hat{\branch} = q_{0} q_{1} \cdots$.
Thus $w$ is accepted by $\aut$ if and only if there exists an $\omega$-branch in $\dagAW{\aut}{w}$ that takes $\accset$-edges infinitely often.

In the remainder of this section, we will introduce the  algorithms for computing the successors of the current states inside different types of SCCs, with the help of run DAGs.
We fix an NBA $\aut = (\states, \init, \trans, \accset)$ and a word $w \in \infwords$.
We let $\states = \setnocond{q_{1}, \dots, q_{n}}$ and apply a total order $\stateOrder$ on $\states$ such that $q_{i} \stateOrder q_{j}$ if $i < j$.
Let  $S_{\ell} \subseteq \states$, $\ell \in \naturals$, be the set of states reached at the level $\ell$ in the run DAG $\dagAW{\aut}{w}$;
we assume that this sequence $S_{0}, \cdots, S_{\ell}, \cdots$ is available as a \emph{global} variable during the computations of every SCC where $S_{0} = \setnocond{\init}$ and $S_{\ell + 1} = \trans(S_{\ell}, \wordletter{w}{\ell})$.

When determinizing the given NBA $\aut$, we classify its SCCs into three types, namely inherently weak SCCs (IWCs), deterministic-accepting SCCs (DACs) and nondeterministic-accepting SCCs (NACs).
We assume that all DACs and NACs are \emph{not} inherently weak, otherwise they will be classified as IWCs.

In our determinization construction, every level in $\dagAW{\aut}{w}$ corresponds to a state in our constructed DELA $\dela$ while reading the $\omega$-word $w$.
Let $m_{\ell}$ be the state of $\dela$ at level $\ell$.
The computation of the successor $m_{\ell + 1}$ of $m_{\ell}$ for the letter $\wordletter{w}{\ell}$ will be divided into the successor computation for states in IWCs, DACs and NACs independently.
Then the successor $m_{\ell + 1}$  is just the Cartesian product of these successors.
In the remainder of this section, we present how to compute the successors for the states in each type of SCCs.

\subsection{Successor Computation inside IWCs}
\label{ssec:determinizationSCCs:IWCs}

As we have seen, $\dagAW{\aut}{w}$ contains all runs of $\aut$ over $w$, including those within DACs and NACs.
Since we want to compute the successor only for IWCs, we focus on the states inside the IWCs and ignore other states in DACs and NACs.
Let $\reachweak$ be the set of states in all IWCs and $\reachweakacc \subseteq \reachweak$ be the set of states in all accepting IWCs.

For the run DAG $\dagAW{\aut}{w}$, we use a pair of sets of states $(P_{\ell}, O_{\ell}) \in 2^{\reachweak} \times 2^{\reachweakacc}$ to represent the set of IWC states reached in $\dagAW{\aut}{w}$ at level $\ell$.
The set $P_{\ell}$ is used to keep track of the states in $\reachweak$ reached at level $\ell$, while $O_{\ell}$, inspired by the breakpoint construction used in~\cite{miyano1984alternating}, keeps only the states reached in $\reachweakacc$, that is, it is used to track the runs that stay in accepting IWCs.
Since by definition each cycle inside an accepting IWC must visit an accepting transition, for each run tracked by $O_{\ell}$ we do not need to remember whether we have taken an accepting transition:
it suffices to know whether the run is still inside some accepting IWC or whether the run has left them.

We now show how to compute the sets $(P_{\ell}, O_{\ell})$ along $w$.
For level $0$, we simply set $P_{0} = \setnocond{\init} \cap \reachweak$ and $O_{0} = \emptyset$.
For the other levels, given $(P_{\ell}, O_{\ell})$ at level $\ell \in \naturals$, the encoding $(P_{\ell + 1}, O_{\ell + 1})$ for the next level $\ell + 1$ is defined as follows:
\begin{itemize}
\item
    $P_{\ell + 1} = S_{\ell + 1} \cap \reachweak$, i.e., $P_{\ell + 1}$ keeps track of the $\reachweak$-states reached at level $\ell + 1$;
\item
    if $O_{\ell} \neq \emptyset$, then $O_{\ell + 1} = \trans(O_{\ell}, \wordletter{w}{\ell}) \cap \reachweakacc$, otherwise $O_{\ell + 1} = P_{\ell + 1} \cap \reachweakacc$.
\end{itemize}
Intuitively, the $O$-set keeps track of the runs that stay in the accepting IWCs.
So if $O_{\ell} \neq \emptyset$, then $O_{\ell + 1}$ maintains the runs remaining in some accepting IWC;
otherwise, $O_{\ell} = \emptyset$ means that at level $\ell$ all runs seen so far in the accepting IWCs have left them, so we can just start to track the new runs that entered the accepting IWCs but were not tracked yet.

\begin{wrapfigure}[9]{r}{24mm}
    \vspace{-9mm}
    \centering
    \resizebox{\linewidth}{!}{
    \includegraphics{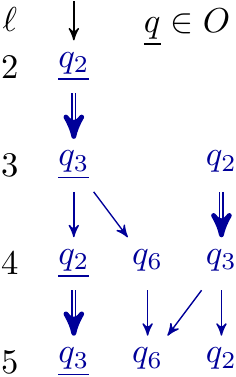}
    }
\end{wrapfigure}

    On the right we show the fragment of the run DAG $\dagAW{\aut}{a^{\omega}}$ for the NBA $\aut$ shown in Figure~\ref{fig:differentSCCs} and its IWCs;
    we have  $\reachweak = \setnocond{q_{2}, q_{3}, q_{6}}$ and $\reachweakacc = \setnocond{q_{2}, q_{3}}$.
    The set $P_{\ell}$ contains all states $q$ at level $\ell$;
    the set $O_{\ell}$ contains the underlined ones.
    As a concrete application of the construction given above, from $P_{3} = \setnocond{q_{2}, q_{3}}$ and $O_{3} = \trans(O_{2}, a) \cap \reachweakacc = \setnocond{q_{3}}$, at level $4$ we get $P_{4} = \setnocond{q_{2}, q_{3}, q_{6}}$ and $O_{4} = \trans(O_{3}, a) \cap \reachweakacc = \setnocond{q_{2}}$.

It is not difficult to see that
checking whether $w$ is accepted reduces to check whether the number of empty $O$-sets is finite.
We assign color $1$ to the transition from $(P_{\ell}, O_{\ell})$ to $(P_{\ell + 1}, O_{\ell+1})$ via $\wordletter{w}{\ell}$ if $O_{\ell} = \emptyset$, otherwise we assign color $2$.
Lemma~\ref{lem:acc-cond-iwcs} formalizes the relation between accepting runs staying in accepting IWCs and the colors we get from our construction.

\begin{restatable}{lemma}{AccIWCs}
\label{lem:acc-cond-iwcs}
    \begin{inparaenum}[(1)]
    \item
    \label{lem:acc-cond-iwcs:correctness}
        There exists an accepting run of $\aut$ over $w$ eventually staying in an accepting IWC if and only if we receive color $1$ finitely many times when constructing the sequence $(P_{0}, O_{0}) \cdots (P_{\ell}, O_{\ell}) \cdots$ while reading $w$.
    \item
    \label{lem:acc-cond-iwcs:complexity}
        The number of possible $(P, O)$ pairs is at most $3^{\size{\reachweak}}$.
    \end{inparaenum}
\end{restatable}
The proof idea is trivial:
an accepting run $\run$ that stays in an accepting IWC will make the $O$-set contain $\run$ forever and we always get color $2$ from some point on.
A possible pair $(P, O)$ can be seen as choosing a state from $\reachweak$, which can be from $\reachweak \setminus P$, $P \cap O$ and $P \setminus O$, respectively.
It thus gives at most $3^{\size{\reachweak}}$ possibilities.
We refer to Appendix~\ref{app:proofs-acc-cond-iwcs} for the detailed proof of Lemma~\ref{lem:acc-cond-iwcs}.

To ease the construction for the whole NBA $\aut$, we make the above computation of successors available as a function $\weaksucc$, which takes as input a pair of sets $(P, O)$ and a letter $a$, and returns the successor $(P', O')$ and the corresponding color $c \in \setnocond{1, 2}$ for the transition $((P, O), a, (P', O'))$.

The construction we gave above works on all IWCs at the same time;
considering IWCs separately does not improve the resulting complexity.
If there are two accepting IWCs with $n_{1}$ and $n_{2}$ states, respectively, then the number of possible $(P, O)$ pairs for the two IWCs is $3^{n_{1}}$ and $3^{n_{2}}$, respectively.
When combining the pairs for each IWC together, the resulting number of pairs in the Cartesian product is $3^{n_{1}} \times 3^{n_{2}} = 3^{n_{1} + n_{2}}$, which is the same as considering them together.
On the other hand, for each accepting IWC, we need to use two colors, so we need $2 \cdot i$ colors in total for $i$ accepting IWCs, instead of just two colors by operating on all IWCs together.
Hence, we prefer to work on all IWCs at once.

\subsection{Successor Computation inside DACs}
\label{ssec:determinizationSCCs:DACs}

In contrast to IWCs, we do not work on all DACs at once but we process each DAC separately.
This is because there may be nondeterminism between DACs: a run in a DAC may branch into multiple runs that jump to different DACs, which requires us to resort to a Safra-Piterman's construction~\cite{safra1988complexity,DBLP:journals/lmcs/Piterman07} when considering all DACs at once.
Working on each DAC separately, instead, allows us to take advantage of the internal determinism:
for a given DAC $\reachdet$, the transition relation $\trans$ inside $\reachdet$, denoted as $\dtrans = (\reachdet \times \alphabet \times \reachdet) \cap \trans$, is now deterministic.

Although every run $\run$ entering $\reachdet$ can have only one successor in $\reachdet$, $\run$ may just leave $\reachdet$ while new runs can enter $\reachdet$, which makes it difficult to check whether there exists an accepting run that remains trapped into $\reachdet$.
In order to identify accepting runs staying in $\reachdet$, we identify the following two rules for distinguishing runs that come to $\reachdet$ by means of \emph{unique} labelling numbers:

\begin{inparaenum}[\bf (1)]
\item
\label{rule:labelling:earlier}
    the runs already in $\reachdet$ have precedence over newly entering runs, thus the latter get assigned a higher number.
    In practice, the labelling keeps track of the relative order of entering $\reachdet$, thus the lower the labelling value is, the earlier the run came to $\reachdet$;
\item
\label{rule:labelling:merge}
    when two runs in $\reachdet$ merge, we only keep the run that came to $\reachdet$ earlier, i.e., the run with lower number.
    If two runs enter $\reachdet$ at the same time, we let them enter according to the total state order $\stateOrder$ for their respective entry states.
\end{inparaenum}

We use a level-labelling function $\labelling_{\ell} \colon \reachdet \to \setnocond{1, \cdots, 2 \cdot \size{\reachdet}} \cup \setnocond{\labellingUndef}$ to encode the set of $\reachdet$-states reached at level $\ell$ of the run DAG $\dagAW{\aut}{w}$.

Here we use $\labelling_{\ell}(q) = \labellingUndef$ to indicate that the state $q \in \reachdet$ is not reached by $\aut$ at level $\ell$.

At level $0$, we set $\labelling_{0}(q) = \labellingUndef$ for every state $q \in \reachdet \setminus \setnocond{\init}$, and $\labelling_{0}(\init) = 1$ if $\init \in \reachdet$.
Note that the SCC that $\init$ resides in can be an IWC, a DAC or a NAC.

For a given level-labelling function $\labelling_{\ell}$, we will make $\setcond{q \in \reachdet}{\labelling_{\ell}(q) \neq \labellingUndef} = S_{\ell} \cap \reachdet$ hold, i.e., tracing correctly the set of $\reachdet$-states reached by $\aut$ at level $\ell$;
we denote the set $\labelling_{\ell}(\reachdet) \setminus \setnocond{\labellingUndef}$ by $\values{\labelling_{\ell}}$, so $\values{\labelling_{\ell}}$ is the set of unique labelling numbers at level $\ell$.
By the construction given below about how to generate $\labelling_{\ell + 1}$ from $\labelling_{\ell}$ on reading $\wordletter{w}{\ell}$, we ensure that $\values{\labelling_{\ell}} \subseteq \setnocond{1, \cdots, 2 \cdot \size{\reachdet}}$ for all $\ell \in \naturals$.

We now present how to compute the successor level-labelling function $\labelling_{\ell + 1}$ of $\labelling_{\ell}$ on letter $\wordletter{w}{\ell}$.
The states reached by $\aut$ at level $\ell + 1$, i.e., $S_{\ell + 1} \cap \reachdet$, may come from two sources:
some state may come from states not in $\reachdet$ via transitions in $\trans \setminus \dtrans$;
some other via $\dtrans$ from  states in $S_{\ell} \cap \reachdet$.
In order to generate $\labelling_{\ell + 1}$, we first compute an intermediate level-labelling function $\labelling'_{\ell + 1}$ as follows.
\begin{enumerate}
\item
\label{def:DAC:labellingprimeDtrans}
    To obey Rule~\eqref{rule:labelling:merge}, for every state $q' \in \dtrans(S_{\ell} \cap \reachdet, \wordletter{w}{\ell})$, we set
    \[
        \labelling'_{\ell + 1}(q') = \min \setcond{ \labelling_{\ell}(q)}{q \in S_{\ell} \cap \reachdet \land \dtrans(q, \wordletter{w}{\ell}) = q'}.
    \]
    That is, when two runs merge, we only keep the run with the lower labelling number, i.e., the run entered in $\reachdet$ earlier.
\item
\label{def:DAC:labellingprimeEntering}
    To respect Rule~\eqref{rule:labelling:earlier}, we set $\labelling'_{\ell + 1}(q') = \size{\reachdet} + i$ for the $i$-th newly entered state $q' \in (S_{\ell + 1} \cap \reachdet) \setminus \dtrans(S_{\ell} \cap \reachdet, \wordletter{w}{\ell})$ and the states $q'$ are ordered by the total order $\stateOrder$ of the states.
    Since every state in $\dtrans(S_{\ell} \cap \reachdet, \wordletter{w}{\ell})$ is on a run that already entered $\reachdet$, its labelling has already been determined by the case~\ref{def:DAC:labellingprimeDtrans}.
\end{enumerate}
It is easy to observe that in order to compute the transition relation between two consecutive levels, we only need to know the labelling at the previous level.
More precisely, we do not have to know the exact labelling numbers, since it suffices to know their relative order.
Therefore, we can compress the level-labelling $\labelling'_{\ell + 1}$ to $\labelling_{\ell + 1}$ as follows.
Let $\orderPosition \colon \values{\labelling'_{\ell + 1}} \to \setnocond{1, \cdots, \size{\values{\labelling'_{\ell + 1}}}}$ be the function that maps each labelling value in $\values{\labelling'_{\ell + 1}}$ to its relative position once the values in $\values{\labelling'_{\ell + 1}}$ have been sorted in ascending order.
For instance, if $\values{\labelling'_{\ell + 1}} = \setnocond{2, 4, 7}$, then $\orderPosition = \setnocond{2 \mapsto 1, 4 \mapsto 2, 7 \mapsto 3}$.
Then we set $\labelling_{\ell + 1}(q) = \orderPosition( \labelling'_{\ell + 1}(q))$ for each $q \in S_{\ell + 1} \cap \reachdet$, and $\labelling_{\ell + 1}(q') = \labellingUndef$ for each $q' \in \reachdet \setminus S_{\ell + 1}$.
In this way, all level-labelling functions $\labelling_{\ell}$ we use are such that $\values{\labelling_{\ell}} \subseteq \setnocond{1, \cdots, \size{\reachdet}}$.

The intuition behind the use of these level-labelling functions is that, if we always see a labelling number $h$ in the intermediate level-labelling $\labelling'_{\ell}$ for all $\ell \geq k$ after some level $k$, we know that there is a run that eventually stays in $\reachdet$ and is eventually always labelled with $h$.
To check whether this run also visits infinitely many accepting transitions, we will color every transition $e = (\labelling_{\ell}, \wordletter{w}{\ell}, \labelling_{\ell + 1}) $.
To decide what color to assign to $e$, we first identify which runs have merged with others or got out of $\reachdet$ (corresponding to \emph{bad} events and \emph{odd} colors) and which runs still continue to stay in $\reachdet$ and take an accepting transition (corresponding to \emph{good} events and \emph{even} colors).

The bad events correspond to the discontinuation of labelling values between $\labelling_{\ell}$ and $\labelling'_{\ell + 1}$, defined as $\parityTranRejValues(e) = \values{\labelling_{\ell}} \setminus \values{\labelling'_{\ell + 1}}$.
Intuitively, if a labelling value $k$ exists in the set $\parityTranRejValues(e)$, then the run $\run$ associated with labelling $k$ merged with a run with lower labelling value $k' < k$, or $\run$ left the DAC $\reachdet$.
The good events correspond to the occurrence of accepting transitions in some runs, whose labelling we collect into

$
    \parityTranAccValues(e) = \setcond{k \in \values{\labelling_{\ell}}}{\exists (q, \wordletter{w}{\ell}, q') \in \accset. \labelling_{\ell}(q) = \labelling'_{\ell + 1}(q') = k \neq \labellingUndef}
$.
In practice, a labelling value $k$ in $\parityTranAccValues(e)$ indicates that we have seen a run with labelling $k$ that visits an accepting transition.
We then let $\parityTranRejValues(e) = \parityTranRejValues(e)\cup \setnocond{\size{\reachdet} + 1}$ and $\parityTranAccValues(e) = \parityTranAccValues(e)\cup \setnocond{\size{\reachdet} + 1}$ where the value $\size{\reachdet} + 1$ is used to indicate that no bad (i.e., no run merged or left the DAC) or no good (i.e., no run took an accepting transition) events happened, respectively.

In order to declare a sequence of labelling functions as accepting, we want the good events to happen infinitely often and bad events to happen only finitely often, when the runs with bad events have a labelling number lower than that of the runs with good events.
So we assign the color $c = \min \setnocond{2 \cdot \min \parityTranRejValues(e) - 1,  2 \cdot \min \parityTranAccValues(e)}$ to the transition $e$.

Since the labelling numbers are in $\setnocond{1, \cdots, \size{\reachdet}}$, we have that $c \in \setnocond{1, \cdots, 2 \cdot \size{\reachdet} + 1}$.
The intuition why we assign colors in this way is given as the proof idea of the following lemma.
\begin{restatable}{lemma}{DACAcceptingRunSize}
\label{lem:acc-run-dacs}
    \begin{inparaenum}[(1)]
    \item
    \label{lem:acc-run-dacs:correctness}
        An accepting run of $\aut$ over $w$ eventually stays in the DAC $\reachdet$ if and only if the minimal color $c$ we receive infinitely often is even.
    \item
    \label{lem:acc-run-dacs:complexity}
        The number of possible labelling functions $\labelling$ is at most $3 \cdot \size{\reachdet}!$.
    \end{inparaenum}
\end{restatable}
The proof idea is as follows:
an accepting run $\run$ on the word $w$ that stays in $\reachdet$ will have stable labelling number, say $k \geq 1$, after some level since the labelling value cannot increase by construction and is finite.
So all runs on $w$ that have labelling values lower than $k$ will not leave $\reachdet$:
if they would leave or just merge with other runs, their labelling value vanishes, so $\orderPosition$ would decrease the value for $\run$.
This implies that the color we receive afterwards infinitely often is either
\begin{inparaenum}[1)]
\item
    an odd color larger than $2k$, due to vanishing runs with value at least $k + 1$ or simply because no bad or good events occur,
    or
\item
    an even color at most $2k$, depending on whether there is some run with value smaller than $\run$ also taking accepting transitions.
\end{inparaenum}
Thus the minimum color occurring infinitely often is even.
The number of labelling functions $\labelling$ is bounded by $\sum_{i = 0}^{\size{\reachdet}} \binom{\size{\reachdet}}{i} \cdot i! \leq 3 \cdot \size{\reachdet}!$.
We refer to Appendix~\ref{app:LDBAacceptingBranchStableLevelCodeterministicDAG} for the detailed proof of Lemma~\ref{lem:acc-run-dacs}.

\begin{wrapfigure}[8]{r}{38mm}
    \vspace{-6mm}
    \centering
    \resizebox{\linewidth}{!}{
    \includegraphics{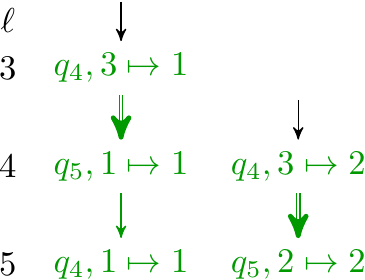}
    }
\end{wrapfigure}

    The fragment of the DAG $\dagAW{\aut}{a^{\omega}}$ shown on the right is relative to the only DAC $\reachdet = \setnocond{q_{4}, q_{5}}$.
    The value of $\labelling'_{\ell}(q)$, $\labelling_{\ell}(q)$ and the corresponding $\orderPosition$ is given by the mapping near each state $q$;
    as a concrete application of the construction given above, consider how to get $\labelling_{4}$ from $\labelling_{3}$, defined as $\labelling_{3}(q_{4}) = 1$ and $\labelling_{3}(q_{5}) = \labellingUndef$:
    since $q_{5} \in \dtrans(S_{3} \cap \reachdet, a)$, according to case~\ref{def:DAC:labellingprimeDtrans}
    we define $\labelling'_{4}(q_{5}) = 1$ because $q_{5} = \dtrans(q_{4}, a)$ and $\labelling_{3}(q_{4}) = 1$;
    since $q_{4} \in (S_{4} \cap \reachdet) \setminus \dtrans(S_{3} \cap \reachdet, a)$, then case~\ref{def:DAC:labellingprimeEntering} applies, so  $\labelling'_{4}(q_{4}) = 3$.
    The function $\orderPosition$ is $\orderPosition = \setn{1 \mapsto 1, 3 \mapsto 2}$, thus we get $\labelling_{4}(q_{4}) = 2$ and $\labelling_{4}(q_{5}) = 1$.
    As bad/good sets for the transition $e = \transition{\labelling_{3}}{a}{\labelling_{4}}$, we have $\parityTranRejValues(e) = \emptyset \cup \setnocond{3}$ while $\parityTranAccValues(e) = \setnocond{1} \cup \setnocond{3}$, so the resulting color is $2$.

Again, we make the above computation of successors available as a function $\detsucc$, which takes as input the DAC $\reachdet$, a labelling $\labelling$ and a letter $a$, and returns the successor labelling $\labelling'$ and the color $c \in \setnocond{1, \cdots, 2 \cdot \size{\reachdet} + 1}$.

\subsection{Successor Computation inside NACs}
\label{ssec:determinizationSCCs:NACs}

The computation of the successor inside a NAC is more involved since runs can branch, so it is more difficult to check whether there exists an accepting run.
To identify accepting runs, researchers usually follow the Safra-Piterman's idea~\cite{safra1988complexity,DBLP:journals/lmcs/Piterman07} to give the runs that take more accepting transitions the precedence over other runs that join them.
We now present how to compute labelling functions encoding this idea for NACs, instead of the whole NBA.
Differently to the previous case about DACs, the labelling functions we use here use lists of numbers, instead of single numbers, to keep track of the branching, merging and new incoming runs.
This can be seen as a generalization of the numbered brackets used in~\cite{DBLP:journals/fuin/Redziejowski12} to represent ordinary Safra-Piterman's trees.
Differently from this construction, in our setting the main challenge we have to consider is how to manage correctly the newly entering runs, which are simply not occurring in~\cite{DBLP:journals/fuin/Redziejowski12} since there the whole NBA is considered.
The fact that runs can merge, instead, is a common aspect, while the fact that a run $\run$ leaves the current NAC can be treated similarly to dying out runs in~\cite{DBLP:journals/fuin/Redziejowski12}.
Below we assume that $\reachnondet$ is a given NAC;
we denote by $\ntrans = (\reachnondet \times \alphabet \times \reachnondet) \cap \trans$ the transition function $\trans$ inside $\reachnondet$.

To manage the branching and merging of runs of $\aut$ over $w$ inside a NAC, and to keep track of the accepting transitions taken so far, we use level-labelling functions as for the DAC case.
For a given NAC $\reachnondet$, the functions we use have lists of natural numbers as codomain;
more precisely, let $\listsnac$ be the set of lists taking value in the set $\setnocond{1, \cdots, 2 \cdot \size{\reachnondet}}$, where a list is a finite sequence of values in ascending order.
Given two lists $\listnac{v_{1}, \cdots, v_{k}}$ and $\listnac{v'_{1}, \cdots, v'_{k'}}$, we say that $\listnac{v_{1}, \cdots, v_{k}}$ is a prefix of $\listnac{v'_{1}, \cdots, v'_{k'}}$ if $1 \leq k \leq k'$ and for each $1 \leq j \leq k$, we have $v_{j} = v'_{j}$.
Note that the empty list \emph{is not} a prefix of any list.
Given two lists $\listnac{v_{1}, \cdots, v_{k}}$ and $\listnac{v'_{1}, \cdots, v'_{k'}}$, we denote by $\listconcat{\listnac{v_{1}, \cdots, v_{k}}}{\listnac{v'_{1}, \cdots, v'_{k'}}}$ their concatenation, that is the list $\listnac{v_{1}, \cdots, v_{k}, v'_{1}, \cdots, v'_{k'}}$.
Moreover, we define a total order on lists as follows:
given two lists $\listnac{v_{1}, \cdots, v_{k}}$ and $\listnac{v'_{1}, \cdots, v'_{k'}}$, we order them by padding the shorter of the two with $\infty$ in the rear, so to make them of the same length, and then by comparing them by the usual lexicographic order.
This means, for instance, that the empty list $\listnac{}$ is the largest list and that $\listnac{1, 3, 5}$ is smaller than $\listnac{1, 3}$ but larger than $\listnac{1, 2}$.
The lists help to keep track of the branching history from their prefixes, such as $[1,2]$ is branched from $[1]$.

As done for DACs, we use a level-labelling function $\labellingtree_{\ell} \colon \reachnondet \to \listsnac$ to encode the set of $\reachnondet$-states reached in the run DAG $\dagAW{\aut}{w}$ at level $\ell$.
We denote by $\values{\labellingtree_{\ell}}$ the set of non-empty lists in the image of $\labellingtree_{\ell}$, that is, $\values{\labellingtree_{\ell}} = \setcond{\labellingtree_{\ell}(q)}{q \in \reachnondet \land \labellingtree_{\ell}(q) \neq \listnac{}}$.
We use the empty list $\listnac{}$ for the states in $\reachnondet$ that do not occur in the vertexes of $\dagAW{\aut}{w}$ at level $\ell$, so $\values{\labellingtree_{\ell}}$ contains only lists associated with states that $\aut$ is currently located at.
Similarly to the other types of SCCs, at level $0$, we set $\labellingtree_{0}(\init) = \listnac{1}$ if $\init \in \reachnondet$, and $\labellingtree_{0}(q) = \listnac{}$ for each state $q \in \reachnondet \setminus \setnocond{\init}$.

To define the transition from $\labellingtree_{\ell}$ to $\labellingtree_{\ell + 1}$ through the letter $\wordletter{w}{\ell}$, we use again an intermediate level-labelling function $\labellingtree'_{\ell + 1}$ that we construct step by step as follows.
We start with $\labellingtree'_{\ell + 1}(q) = \listnac{}$ for each $q \in \reachnondet$ and with the set of unused numbers $U = \setcond{ u \geq 1 }{u \notin \values{\labellingtree_{\ell}}}$, i.e., the numbers not used in $\values{\labellingtree_{\ell}}$.

\begin{enumerate}
\item
\label{def:NAC:labellingprimeNtrans}
	For every state $q' \in \ntrans(S_{\ell} \cap \reachnondet, \wordletter{w}{\ell})$, let $P_{q'} = \setcond{q \in S_{\ell} \cap \reachnondet}{(q, \wordletter{w}{\ell}, q') \in \ntrans}$ be the set of currently reached predecessors of $q'$, and $C_{q'} = \emptyset$.
	For each $q \in P_{q'}$, if $(q, \wordletter{w}{\ell}, q') \in \accset$, then we add $\listconcat{\labellingtree_{\ell}(q)}{\listnac{u}}$ to $C_{q'}$, where $u = \min U$, and we remove $u$ from $U$, so that each number in $U$ is used only once;
	otherwise, for $(q, \wordletter{w}{\ell}, q') \in \ntrans \setminus \accset$, we add $\labellingtree_{\ell}(q)$ to $C_{q'}$.
	Lastly, we set $\labellingtree'_{\ell + 1}(q') = \min C_{q'}$, where the minimum is taken according to the list order.
	
	Intuitively, if a run $\run$ can branch into two kinds of runs, some via accepting transitions and some others via nonaccepting transitions at level $\ell + 1$, then we let those from nonaccepting transitions inherit the labelling from $\run$, i.e., $\labellingtree_{\ell}(\wordletter{\run}{\ell})$;
	for the runs taking accepting transitions we create a new labelling $\listconcat{\labellingtree_{\ell}(\wordletter{\run}{\ell})}{\listnac{u}}$.
	In this way, the latter get precedence over the former.
	Moreover, if a run $\run$ has received multiple labelling values, collected in $C_{\wordletter{\run}{\ell + 1}}$, then it will keep the smallest one, by $\labellingtree'_{\ell + 1}(\wordletter{\run}{\ell + 1}) = \min C_{\wordletter{\run}{\ell + 1}}$.
\item
\label{def:NAC:labellingprimeEntering}
	For each state $q' \in (S_{\ell + 1} \cap \reachnondet) \setminus \ntrans(S_{\ell} \cap \reachnondet, \wordletter{w}{\ell})$ taken according to the state order $\stateOrder$, we first set $\labellingtree'_{\ell + 1} (q') = \listnac{u}$, where $u = \min U$, and then we remove $u$ from $U$, so we do not reuse the same values.
	That is, we give the newly entered runs lower precedence than those already in $\reachnondet$, by means of the larger list $\listnac{u}$.
\end{enumerate}
We now need to prune the lists in $\values{\labellingtree'_{\ell + 1}}$ and recognize good and bad events.
Similarly to DACs, a bad event means that a run has left $\reachnondet$ or has been merged with runs with smaller labelling, which is indicated by a discontinuation of a labelling between $\values{\labellingtree_{\ell}}$ and $\values{\labellingtree'_{\ell + 1}}$.
For the transition $e = (\labellingtree_{\ell}, \wordletter{w}{\ell}, \labellingtree_{\ell +1})$ we are constructing, to recognize bad events, we put into the set $\parityTranRejValues(e)$ the number $\size{\reachnondet} + 1$ and all numbers in $\values{\labellingtree_{\ell}}$ that have disappeared in $\values{\labellingtree'_{\ell + 1}}$, that is, $\parityTranRejValues(e) = \setnocond{\size{\reachnondet} + 1} \cup \setcond{v \in \naturals}{\text{$v$ occurs in $\values{\labellingtree_{\ell}} $ but not in $ \values{\labellingtree'_{\ell + 1}}$}}$.

Differently from the good events for DACs, which require to visit an accepting transition, we need all runs branched from a run to visit an accepting transition, which is indicated by the fact that there are no states labelled by $\labellingtree'_{\ell + 1}$ with some list $l \in \values{\labellingtree_{\ell}}$ but there are extensions of $l$ associated with some state.
To recognize good events, let $\parityTranAccValues(e) = \setnocond{\size{\reachnondet} + 1}$ and $\labellingtree''_{\ell + 1}$ be another intermediate labelling function.
For each $q' \in S_{\ell + 1} \cap \reachnondet$, consider the list $\labellingtree'_{\ell + 1}(q')$:
if for each prefix $\listnac{v_{1}, \cdots v_{k}}$ of $\labellingtree'_{\ell + 1}(q')$ we have $\listnac{v_{1}, \cdots v_{k}} \in \values{\labellingtree'_{\ell + 1}}$, then we set $\labellingtree''_{\ell + 1}(q') = \labellingtree'_{\ell + 1}(q')$.
Otherwise, let $\listnac{v_{1}, \cdots v_{\bar{k}}} \notin \values{\labellingtree'_{\ell + 1}}$ be the shortest prefix of $\labellingtree'_{\ell + 1}(q')$ not in $\values{\labellingtree'_{\ell + 1}}$;
we set $\labellingtree''_{\ell + 1}(q') = \listnac{v_{1}, \cdots v_{\bar{k}}}$ and add $v_{\bar{k}}$ to $\parityTranAccValues(e)$.
Setting $\labellingtree''_{\ell + 1}(q') = \listnac{v_{1}, \cdots v_{\bar{k}}}$ in fact corresponds, in the Safra's construction~\cite{safra1988complexity}, to the removal of all children of a node $\mathfrak{N}$ for which the union of the states in the children is equal to the states in $\mathfrak{N}$.
Lastly, similarly to the DAC case, we set $\labellingtree_{\ell + 1}(q) = \orderPosition(\labellingtree''_{\ell + 1}(q))$ for each $q \in S_{\ell + 1} \cap \reachnondet$ and $\labellingtree_{\ell + 1}(q') = \listnac{}$ for each $q' \in \reachnondet \setminus S_{\ell + 1}$, where $\orderPosition(\listnac{v_{1}, \cdots, v_{k}}) = \listnac{\orderPosition(v_{1}), \cdots, \orderPosition(v_{k})}$.
Regarding the color to assign to the transition $e$, we just assign the color $c = \min
\setnocond{2 \cdot \min \parityTranAccValues(e), 2 \cdot \min \parityTranRejValues(e) - 1}$.

\begin{restatable}{lemma}{nacAccCond}
\label{lem:nac-acc-cond}
    \begin{inparaenum}[(1)]
    \item
    \label{lem:nac-acc-cond:correctness}
        An accepting run of $\aut$ over $w$ eventually stays in the NAC $\reachnondet$ if and only if the minimal color $c$ we receive infinitely often is even.
    \item
    \label{lem:nac-acc-cond:complexity}
        The number of possible labelling functions $\labellingtree$ is at most $2 \cdot (\size{\reachnondet}!)^{2}$.
    \end{inparaenum}
\end{restatable}

Similarly to DACs, also for NACs we have handled each NAC independently.
The reason for this is that this potentially reduces the complexity of the single cases:
assume that we have two NACs $\reachnondet_{1}$ and $\reachnondet_{2}$.
If we apply the Safra-Piterman's construction directly to $\reachnondet_{1} \cup \reachnondet_{2}$, we might incur in  the worst-case complexity $2 \cdot ((\size{\reachnondet_{1}} + \size{\reachnondet_{2}})!)^{2}$, as mentioned in the introduction.
However, if we determinize them separately, then the worst complexity for each NAC $\reachnondet_{i}$ is $2 \cdot (\reachnondet_{i}!)^{2}$, for an overall  $4 \cdot (\size{\reachnondet_{1}}! \cdot \size{\reachnondet_{2}}!)^{2}$, much smaller than $2 \cdot ((\size{\reachnondet_{1}} + \size{\reachnondet_{2}})!)^{2}$.

As usual, we make the above construction available as a function $\nondetsucc$, which takes as input the NAC $\reachnondet$, a labelling $\labellingtree$ and a letter $a$, and returns the successor labelling $\labellingtree'$ and the corresponding color $c \in \setnocond{1, \cdots, 2 \cdot \size{\reachnondet} + 1}$.

\begin{wrapfigure}[7]{r}{30mm}
    \vspace{-8mm}
    \centering
    \resizebox{\linewidth}{!}{
    \includegraphics{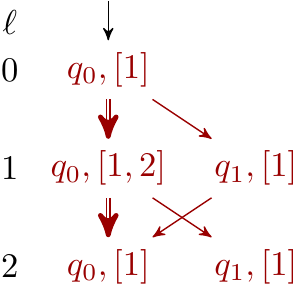}
    }
\end{wrapfigure}

    Similarly to the constructions for other SCCs, we show on the right the fragment of run DAG $\dagAW{\aut}{a^{\omega}}$ for the NAC $\reachnondet = \setnocond{q_{0}, q_{1}}$, with $q_{0} \stateOrder q_{1}$.
    The construction of $\labellingtree_{1}$ is easy, so consider its $a$-successor $\labellingtree_{2}$:
    we start with $U = \setnocond{3, 4,\cdots}$;
    for $q_{0}$, we have $P_{q_{0}} = \setnocond{q_{0}, q_{1}}$ and $C_{q_{0}} = \setnocond{\listnac{1, 2, 3}, \listnac{1}}$, hence $\labellingtree'_{2}(q_{0}) = \listnac{1, 2, 3}$.
    For $q_{1}$, we get $P_{q_{1}} = \setnocond{q_{0}}$ and $C_{q_{1}} = \setnocond{\listnac{1, 2}}$, so $\labellingtree'_{2}(q_{1}) = \listnac{1, 2}$.
    Thus, for $e = (\labellingtree_{1}, \wordletter{w}{1}, \labellingtree_{2})$, we have $\parityTranRejValues(e) = \setnocond{3}$ while $\parityTranAccValues(e) = \setnocond{1, 3}$, since both lists in $\values{\labellingtree'_{2}} = \setnocond{\listnac{1,2}, \listnac{1,2,3}}$ are missing the prefix $\listnac{1}$, so we get  $\labellingtree_{2}(q_{0}) = \labellingtree_{2}(q_{1}) = \listnac{1}$ and color $c = 2$.

\section{Determinization of NBAs to DELAs}
\label{sec:LDBADeterminizationToParity}

In this section, we fix an NBA $\aut = (\states, \init, \trans, \accset)$ with $n = \size{\states}$ states and we show how to construct an equivalent DELA $\dela = (\delaStates, \delaInitial, \delaTrans,  \delaColor, \delaParity, \delaAcc)$, by using the algorithms developed in the previous section.
We assume that $\aut$ has $\setnocond{\reachdet^{1}, \cdots, \reachdet^{d}}$ as set of DACs and $\setnocond{\reachnondet^{1}, \cdots, \reachnondet^{k}}$ as set of NACs.

When computing the successor for each type of SCCs while reading a word $w$, we just need to know the set $S_{\ell}$ of states reached at the current level $\ell$ and the letter $a \in \alphabet$ to read.
We can ignore the actual level $\ell$, since if $S_{\ell} = S_{\ell'}$, then their successors under the same letter will be the same.
As mentioned before, every state of $\dela$ corresponds to a level of $\dagAW{\aut}{w}$.
We call a state of $\dela$ a \emph{macrostate} and a run of $\dela$ a \emph{macrorun}, to distinguish them from those of $\aut$.

\paragraph*{\bf Macrostates $\delaStates$.}
Each macrostate consists of the pair $(P, O)$ for encoding the states in IWCs, a labelling function $\labelling^{i} \colon \reachdet^{i} \to \setnocond{1, \cdots, \size{\reachdet^{i}}} \cup \setnocond{\labellingUndef}$ for the states of each DAC $\reachdet^{i}$ and a labelling function $\labellingtree^{j} \colon \reachnondet^{j} \to \listsnac[\reachnondet^{j}]$ for each NAC $\reachnondet^{j}$, without the explicit level number.
The initial macrostate $\delaInitial$ of $\dela$ is the encoding of level $0$, defined as the set $\setnocond{(P_{0}, O_{0})} \cup \setcond{\labelling^{i}_{0}}{\text{$\reachdet^{i}$ is a DAC}} \cup \setcond{\labellingtree^{j}_{0}}{\text{$\reachnondet^{j}$ is a NAC}}$, where each encoding for the different types of SCCs is the one for level $0$.

We note that $\init$ must be present in one type of SCCs.
In particular, if $\init$ is a transient state, then $\setnocond{\init}$ is classified as an IWC.

\paragraph*{\bf Transition function $\delaTrans$.}
Let $m$ be the current macrostate in $\delaStates$ and $a \in \alphabet$ be the letter to read.
Then we define $m' = \delaTrans(m, a)$ as follows.
\begin{enumerate}[(i)]
\item
\label{def:LDBAparityDeterminization:transition:iwc}
    For $(P_{m}, O_{m}) \in m$, we set $(P_{m'}, O_{m'}) = \weaksucc((P_{m}, O_{m}), a)$ in $m'$.
\item
\label{def:LDBAparityDeterminization:transition:dac}
    For $\labelling^{i}_{m} \in m$ relative to the DAC $\reachdet^{i}$, we set  $\labelling^{i}_{m'} = \detsucc(\reachdet^{i}, \labelling^{i}_{m}, a)$ in $m'$.
\item
\label{def:LDBAparityDeterminization:transition:nac}
    For $\labellingtree^{j}_{m} \in m$ from the NAC $\reachnondet^{j}$, we set $\labellingtree^{j}_{m'} = \nondetsucc(\reachnondet^{j}, \labellingtree^{j}_{m}, a)$ in $m'$.
\end{enumerate}
Note that the set $S$ of the current states of $\aut$ used by the different successor functions is implicitly given by the sets $P$, $\setcond{q \in \reachdet^{i}}{\labelling^{i}(q) \neq \labellingUndef}$ for each DAC $\reachdet^{i}$ and $\setcond{q \in \reachnondet^{j}}{\labellingtree^{j}(q) \neq \listnac{}}$ for each NAC $\reachnondet^{j}$ in the current macrostate $m$.

\paragraph*{\bf Color set $\delaColor$ and coloring function $\delaParity$.}

From the constructions given in Section~\ref{sec:determinizationSCCs}, we have two colors from the IWCs, $2 \cdot \size{\reachdet^{i}} + 1$ colors for each DAC $\reachdet^{i}$, and $2 \cdot \size{\reachnondet^{j}} + 1$ colors for each NAC $\reachnondet^{j}$, yielding a total of at most $3 \cdot \size{\states}$ colors.
Thus we set $\delaColor = \setnocond{0, 1, \cdots, 3 \cdot \size{\states}}$ with color $0$ not being actually used.

Regarding the color to assign to each transition, we need to ensure that the colors returned by the single SCCs are treated separately, so we transpose them.
For a transition $e = (m, a, m') \in \delaTrans$, we define the coloring function $\delaParity$ as follows.
\begin{itemize}
\item
    If we receive color $1$ for the transition $((P_{m}, O_{m}), a, (P_{m'}, O_{m'}))$, then we put $1 \in \delaParity(e)$.
    Intuitively, every time we see an empty $O$-set along reading an $\omega$-word $w$ in the IWCs, we put the color $1$ on the transition $(m, a, m')$.
\item
    For each DAC $\reachdet^{i}$, we transpose its colors after the colors for the IWCs and the other DACs with smaller index.
    So we set the base number for the colors of the DAC $\reachdet^{i}$ to be $\basenumber_{i} = 2 + \sum_{1 \leq h < i} (2 \cdot \size{\reachdet^{h}} + 1)$, i.e., the number of colors already being used.
    Then, if we receive the color $c$ for the transition $(\labelling^{i}_{m}, a, \labelling^{i}_{m'})$ from $\detsucc$, we put $c + \basenumber_{i} \in \delaParity(e)$.
\item
    We follow the same approach for the NAC $\reachnondet^{j}$:
    we set its base number to be $\basenumber_{j} = 2 + \sum_{1 \leq h \leq d} (2 \cdot \size{\reachdet^{h}} + 1) + \sum_{1 \leq h < j} (2 \cdot \size{\reachnondet^{h}} + 1)$.
    Then, if we receive the color $c$ for the transition $(\labellingtree^{j}_{m}, a, \labellingtree^{j}_{m'})$ from $\nondetsucc$, we put $c + \basenumber_{j} \in \delaParity(e)$.
\end{itemize}
Intuitively, we make the colors returned for each SCC not overlap with those of other SCCs without changing their relative order.
In this way, we can still independently check whether there exists an accepting run staying in an SCC.

\paragraph*{\bf Acceptance formula $\delaAcc$.}
We now define the acceptance $\delaAcc$, which is basically the \emph{disjunction} of the acceptance formula for each different types of SCCs, after transposing them.
Regarding the IWCs, we trivially define  $\delaAcc_{\reachweak} = \accfin{1}$, since this is the acceptance formula for IWCs;
as said before, color $0$ is not used.

For DACs and NACs, the definition is more involved.
For instance, regarding the DAC $\reachdet^{i}$, we know that all returned colors are inside $\setnocond{1, \cdots, 2 \cdot \size{\reachdet^i} + 1}$.
According to Lemma~\ref{lem:acc-run-dacs}, an accepting run eventually stays in $\reachdet^i$ if and only if the minimum color that we receive infinitely often is even.
Thus, the acceptance formula for the above lemma is
$\mathsf{parity}(\size{\reachdet^{i}}) = \bigvee_{c = 1}^{\size{\reachdet^{i}}} (\bigwedge_{j=1}^{c} \accfin{2j - 1} \land \accinf{2c})$.
Let $\basenumber_{i} = 2 + \sum_{h < i} (2 \cdot \size{\reachdet_{h}} + 1)$ be the base number for the colors of $\reachdet^{i}$, which is also the number of colors already used by IWCs and the DACs $\reachdet^{h}$ with $h < i$.
Since we have added the base number $\basenumber^{i}$ to every color of $\reachdet^{i}$, we then have the acceptance formula
$\delaAcc_{\reachdet^{i}} = \bigvee_{c = 1}^{\size{\reachdet^{i}}} (\bigwedge_{j=1}^{c} \accfin{2j - 1 + \basenumber_{i}} \land \accinf{2c + \basenumber_{i}})$.

For each NAC $\reachnondet^{j}$, the colors we receive are in $\setnocond{1, \cdots, 2 \cdot \size{\reachnondet^{j}} + 1}$.
Let $\basenumber_{j} = 2 + \sum_{1 \leq h \leq d}(2 \cdot \size{\reachdet^{h}} + 1) + \sum_{h < j} (2 \cdot \size{\reachnondet^{j}} + 1)$ be the base number for $\reachnondet^{j}$.
Similarly to the DAC case, for each NAC $\reachnondet^{j}$, we let
$\delaAcc_{\reachnondet^{j}} = \bigvee_{c = 1}^{\size{\reachnondet^{j}}} (\bigwedge_{i = 1}^{c} \accfin{2i - 1 + \basenumber_{j}} \land \accinf{2c + \basenumber_{j}})$.

The acceptance formula for $\dela$ is $\delaAcc = \delaAcc_{\reachweak} \lor \bigvee_{i = 1}^{d} \delaAcc_{\reachdet^{i}} \lor \bigvee_{j = 1}^{k} \delaAcc_{\reachnondet^{j}}$.

Consider again the NBA $\aut$ given in Figure~\ref{fig:differentSCCs} and its various SCCs.
As acceptance formula for the constructed DELA, it is the disjunction of the formulas
$\delaAcc_{\reachweak} = \accfin{1}$;
$\delaAcc_{\reachdet} = \bigvee_{c = 1}^{2} (\bigwedge_{j=1}^{c} \accfin{2j - 1 + 2} \land \accinf{2c + 2})$, since the base number for $\reachdet$ is $2$;
and
$\delaAcc_{\reachnondet} = \bigvee_{c = 1}^{2} (\bigwedge_{i = 1}^{c} \accfin{2i - 1 + 7} \land \accinf{2c + 7})$, since $7$ is the base number for $\reachnondet$.

The construction given in this section is correct, as stated by Theorem~\ref{thm:LDBAlanguageSizeParityAutomaton}.

\begin{restatable}{theorem}{LDBAlanguageSizeParityAutomaton}
\label{thm:LDBAlanguageSizeParityAutomaton}
    Given an NBA $\aut$ with $n = \size{\states}$ states, let $\dela$ be the DELA constructed by our method.
    Then
    \begin{inparaenum}[(1)]
    \item
    \label{thm:LDBAlanguageSizeParityAutomaton:correctness}
        $\lang{\dela} = \lang{\aut}$ and
    \item
    \label{thm:LDBAlanguageSizeParityAutomaton:complexity}
        $\dela$ has at most $3^{\size{\reachweak}} \cdot \Big(\prod_{i=1}^{d} 3 \cdot \size{\reachdet^{i}}! \Big) \cdot \Big( \prod_{j=1}^{k} 2 \cdot (\size{\reachnondet^{i}}!)^{2}\Big)$ macrostates and $3 n + 1$ colors.
    \end{inparaenum}
\end{restatable}

Obviously, if $d = k = 0$, $\aut$ is a weak BA~\cite{DBLP:journals/tcs/MullerSS92}.
If $k = 0$, $\aut$ is an elevator BA, a new class of BAs recently introduced in~\cite{DBLP:conf/tacas/HavlenaLS22} which have only IWCs and DACs, a strict superset of semi-deterministic BAs (SDBAs)~\cite{CourcoubetisY95}.
SDBAs will behave \emph{deterministically} after seeing acceptance transitions.
An elevator BA that is not an SDBA can be obtained from the NBA $\aut$ shown in Figure~\ref{fig:differentSCCs} by setting $q_{2}$ as initial state and by removing all states and transitions relative to the NAC.

It is known that the lower bound for determinizing SDBAs is $n!$~\cite{DBLP:conf/tacas/EsparzaKRS17,DBLP:conf/fsttcs/Loding99}.
Then the determinization complexity of weak BAs and elevator BAs can be easily improved exponentially as follows.
\begin{restatable}{corollary}{worstCaseDeterminizationSubclass}
\label{coro:worst-case-subclasses-determinization}
    \begin{inparaenum}[(1)]
    \item
    \label{coro:worst-case-subclasses-determinization:weak}
        Given a weak \buchi automaton $\aut$ with $n = \size{\states}$ states, the DELA constructed by our algorithm has at most $3^{n}$ macrostates.
    \item
    \label{coro:worst-case-subclasses-determinization:elevator}
        Given an elevator \buchi automaton $\aut$ with $n = \size{\states}$ states, our algorithm constructs a DELA with $\bigM(n!)$ macrostates;
        it is asymptotically optimal.
    \end{inparaenum}
\end{restatable}
The upper bound for determinizing weak BAs is already known~\cite{DBLP:conf/cade/BoigelotJW01}.
Elevator BAs are, to the best of our knowledge, the \emph{largest} subclass of NBAs known so far to have determinization complexity $\bigM(n!)$.

The acceptance formula for an SCC can be seen as a parity acceptance formula with colors being shifted to different ranges.
A parity automaton can be converted into a Rabin one without blow-up of states and transitions~\cite{DBLP:conf/dagstuhl/Farwer01}.
Since $\delaAcc$ is a disjunction of parity acceptance formulas, Theorem~\ref{thm:dela-to-dra} then follows.
\begin{restatable}{theorem}{DELAtoDRA}
\label{thm:dela-to-dra}
    Let $\dela$ be the constructed DELA for the given NBA $\aut$.
    Then $\dela$ can be converted into a DRA $\dra$ without blow-up of states and transitions.
\end{restatable}

\paragraph{\bf Translation to deterministic Parity automata (DPAs).} We note that there is an \emph{optimal} translation from a DRA to a DPA described in\cite{DBLP:conf/icalp/CasaresCF21}, implemented in \spot via the function $\mathsf{acd\_transform}$~\cite{DBLP:conf/tacas/Casares22}.

\section{Empirical Evaluation}
\label{sec:experiments}

To analyze the effectiveness of our Divide-and-Conquer determinization construction proposed in Section~\ref{sec:determinizationSCCs}, we implemented it in our tool \cola, which is built on top of \spot~\cite{duret2016spot}.
The source code of \cola is publicly available from \url{https://github.com/liyong31/COLA}.
We compared \cola with the official versions of \spot~\cite{duret2016spot} (2.10.2) and \owl~\cite{DBLP:conf/atva/KretinskyMS18} (21.0).
\spot implements the algorithm described in~\cite{DBLP:journals/fuin/Redziejowski12}, a variant of~\cite{DBLP:journals/lmcs/Piterman07} for transition-based NBAs, while \owl implements the algorithms described in~\cite{DBLP:conf/icalp/LodingP19,DBLP:conf/atva/LodingP19}, both constructing DPAs as result.
To make the comparison fair, we let all tools generate DPAs, so we used the command \texttt{autfilt --deterministic --parity=min\textbackslash{} even -F file.hoa} to call \spot and  \texttt{owl nbadet -i file.hoa} to call \owl.
Recall that we use the function $\mathsf{acd\_transform}$~\cite{DBLP:conf/tacas/Casares22} from \spot for obtaining DPAs from our DRAs.
The tools above also implement optimizations for reducing the size of the output DPA, like simulation and state merging~\cite{DBLP:conf/atva/LodingP19}, or stutter invariance~\cite{DBLP:conf/wia/KleinB07} (except for \owl);
we use the default settings for all tools.
We performed our experiments on a desktop machine equipped with 16GB of RAM and a 3.6 GHz Intel Core i7-4790 CPU.
We used \benchexec\footnote{\url{https://github.com/sosy-lab/benchexec/}}~\cite{DBLP:journals/sttt/BeyerLW19} to trace and constrain the tools' executions:
we allowed each execution to use a single core and 12 GB of memory, and imposed a timeout of 10 minutes.
We used \spot to verify the results generated by three tools and found only outputs equivalent to the inputs.

As benchmarks, we considered all NBAs in the HOA format~\cite{Babiak15} available in the \automatabenchmarks repository\footnote{\url{https://github.com/ondrik/automata-benchmarks/}}.
We have pre-filtered them with \autfilt to exclude all deterministic cases and to have nondeterministic BAs, obtaining in total 15,913 automata coming from different sources in literature.

\begin{figure}[t!]
    \centering
    \resizebox{\linewidth}{!}{
    \includegraphics{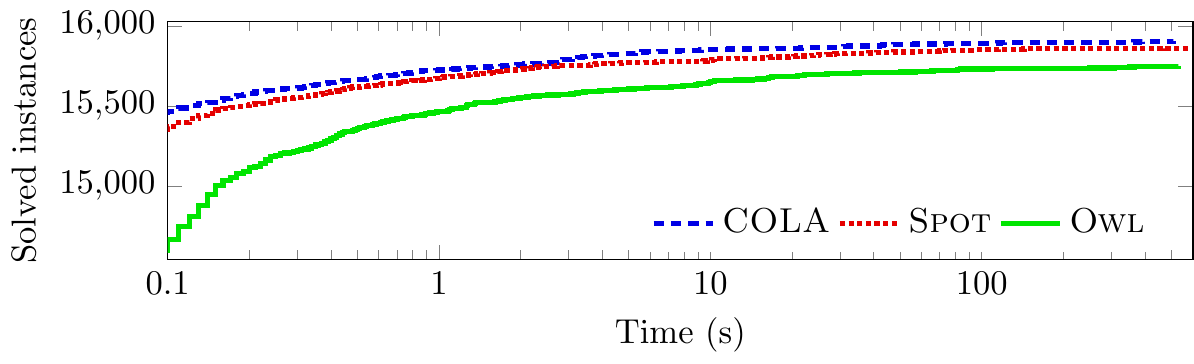}
    }
    \caption{The cactus plot for the determinization of NBAs from \automatabenchmarks.}
    \label{fig:experimentsAutomataParityCactusPlottime}
\end{figure}
In Figure~\ref{fig:experimentsAutomataParityCactusPlottime} we show a cactus plot reporting how many input automata have been determinized by each tool, over time.
As we can see, \cola works better than \spot, with \cola solving in total $15,903$ cases and \spot $15,862$ cases, with \owl solving in total $15,749$ cases and taking more time to solve as many instances as \cola and \spot.
From the plot given in Figure~\ref{fig:experimentsAutomataParityCactusPlottime} we see that \cola is already very competitive with respect to its performance.

\begin{figure}[t!]
    \centering
    \resizebox{\linewidth}{!}{
    \includegraphics{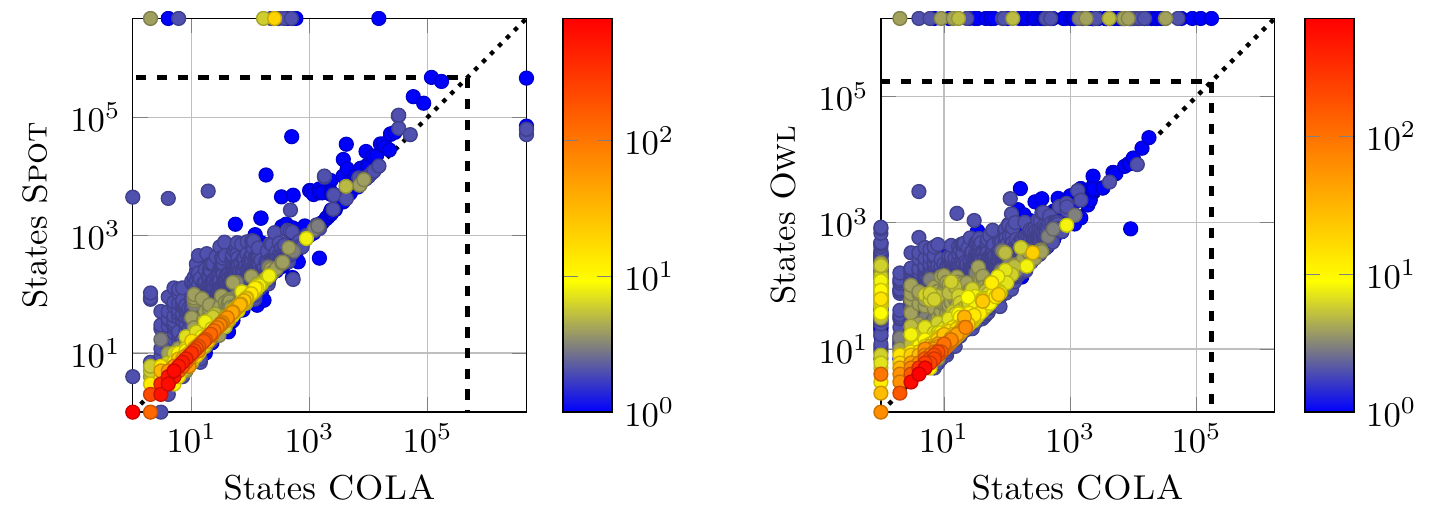}
    }
    \caption{States comparison for the determinization of NBAs from \automatabenchmarks.}
    \label{fig:experimentsAutomataParityStates}
\end{figure}
In Figure~\ref{fig:experimentsAutomataParityStates} we show the number of states of the generated DPAs.
In the plot we indicate with the bold dashed line the maximum number of states of the automata produced by either of the two tools, and we place a mark on the upper or right border of the plot to indicate that one tool has generated an automaton with that size while the other tool just failed.
The color of each mark represents how many instances have been mapped to the corresponding point.
As the plots show, \spot and \cola generate automata with similar size, with \cola being more likely to generate smaller automata, in particular for larger outputs.
\owl, instead, very frequently generates automata larger than \cola.
In fact, on the 15,710 cases solved by all tools, on average \cola generated 44 states, \spot 65, and \owl 87.
If we compare \cola with just one tool at a time, on the 15,854 cases solved by both \cola and \spot, we have 125 states for \cola and 246 for \spot;
on the 15,749 cases solved by both \cola and \owl, we have 45 states for \cola and 88 for \owl.
A similar situation occurs for the number of transitions, so we omit it.

\begin{figure}[t!]
    \centering
    \resizebox{\linewidth}{!}{
    \includegraphics{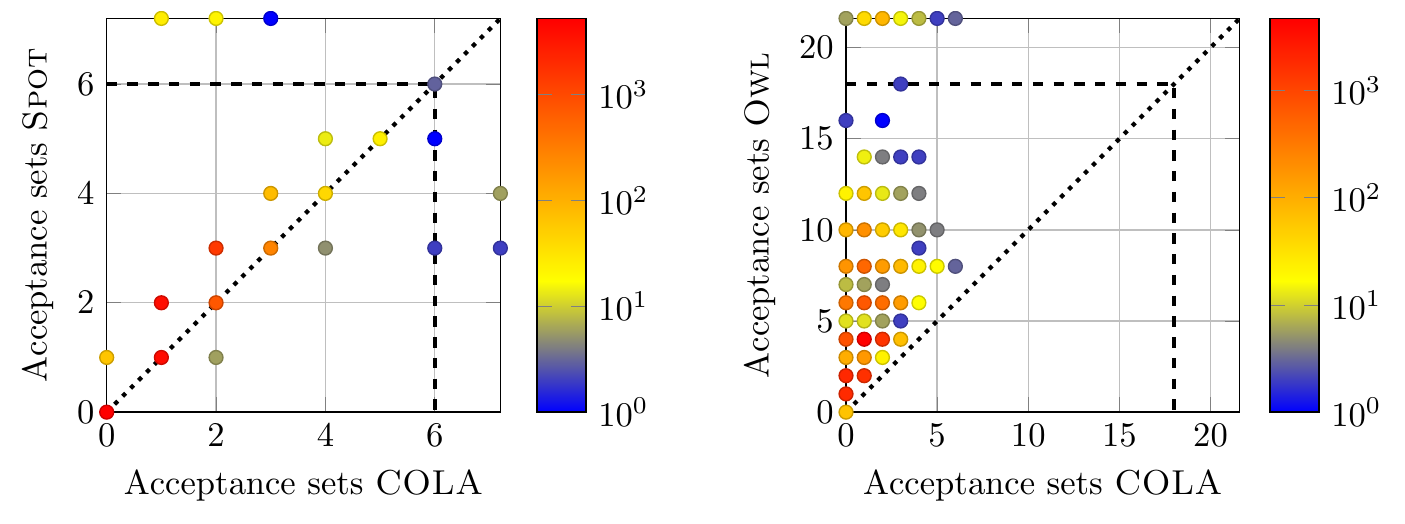}
    }
    \caption{Acceptance sets comparison for the determinization of NBAs from \automatabenchmarks.}
    \label{fig:experimentsAutomataParityAcceptanceSets}
\end{figure}
Lastly, in Figure~\ref{fig:experimentsAutomataParityAcceptanceSets} we compare the number of acceptance sets (i.e., the colors in Definition~\ref{def:automaton}) of the generated DPAs;
more precisely, we consider the integer value occurring in the mandatory \texttt{Acceptance: INT acceptance-cond} header item of the HOA format~\cite{Babiak15}, which can be $0$ for the automata with all or none accepting transitions.
From the plots we can see that \cola generates more frequently DPAs with a number of colors that is no more than the number used by \spot, as indicated by the yellow/red marks on (10,394 cases) or above (5,495 cases) the diagonal.
Only in very few cases \cola generates DPAs with more colors than \spot (22 cases), as indicated by the few blue/greenish marks below the diagonal.
Regarding \owl, however, from the plot we can clearly see that \cola uses almost always (15,840 cases) fewer colors than \owl;
the only exception is for the mark at $(0,0)$ representing 63 cases.

\begin{table}[t]
\setlength{\tabcolsep}{5pt}
    \caption{Pearson correlation coefficients for the \automatabenchmarks experiments.}
    \label{tab:experimentsAutomataPearson}

    \centering
    \begin{tabular}{c|ccc}
        & \# input states & \# input SCCs & average SCC size \\
        \hline
        runtime & 0.77 & 0.62 & -0.01\\
        output states & 0.41 & 0.17 & 0.05 \\
    \end{tabular}
\end{table}
The number and sizes of SCCs influence the performance of \cola, so we provide some statistics about the correlation between these and the runtime and size of the generated DPA.
By combining the execution statistics with the input SCCs and states, we get the Pearson correlation coefficients shown in Table~\ref{tab:experimentsAutomataPearson}.
Here the larger the number in a cell is, the stronger the positive correlation between the element that the row and the column represent.
From these coefficients we can say that there is a quite strong positive correlation between the number of states and of SCCs and the running time, but not for the average SCC size; regarding the output states, the situation is similar but much weaker.

We also considered a second set of benchmarks -- 644 NBAs generated by \spot's \texttt{ltl2tgba} on the LTL formulas considered in~\cite{DBLP:conf/atva/KretinskyMS18}, as available in the \owl's repository at \url{https://gitlab.lrz.de/i7/owl}.
The outcomes for these benchmarks are similar, but a bit better for \cola, to the ones for \automatabenchmarks, so we do not present them in detail.
In Appendix~\ref{app:plotsExperiments} we provide additional plots for the \automatabenchmarks benchmarks as well as the ones for these 644 NBAs.

\section{Related Work}
\label{sec:related-work}

To the best of our knowledge, our determinization construction is the \emph{first} algorithm that determinizes SCCs \emph{independently} while taking advantage of different structures of SCCs, which is the main difference between our algorithm and existing works.
We illustrate other minor differences below.

Different types of SCCs, like DACs and IWCs, are also taken with special care in~\cite{DBLP:conf/atva/LodingP19} as in our work, modulo the handling details.
However, the work~\cite{DBLP:conf/atva/LodingP19} does not treat them independently as the labelling numbers in those SCCs still have relative order with those in other SCCs.
Thus their algorithm can be exponentially worse than ours (cf. Theorem~\ref{thm:better-complexity-nbas}) and performs not as well as ours in practice;
see the comparison with \owl in Section~\ref{sec:experiments}.
The determinization algorithm given in~\cite{DBLP:conf/tacas/EsparzaKRS17} for SDBAs is a special case of the one presented in~\cite{DBLP:journals/fuin/Redziejowski12} for NBAs, which gives precedence to the deterministic runs seeing accepting transitions earlier, while we give precedence to runs that enter DACs earlier.
More importantly, the algorithm from~\cite{DBLP:conf/tacas/EsparzaKRS17} does not work when there is nondeterminism between DACs, while our algorithm overcomes this by considering DACs separately and by ignoring runs going to other SCCs.

Current works for determinization of general NBAs, such as~\cite{DBLP:journals/fuin/Redziejowski12,safra1988complexity,DBLP:journals/iandc/FogartyKVW15,DBLP:conf/icalp/LodingP19,kahler2008complementation,Schewe09} can all be interpreted as different flavours of the Safra-Piterman based algorithm.
Our determinization of NACs is also based on Safra-trees, except that we may have newly arriving states from other SCCs while other works only need to consider the successors from the current states in the Safra-tree.
The modular approach for determinizing \buchi automata given in~\cite{DBLP:conf/concur/FismanL15} builds on reduced split trees~\cite{kahler2008complementation} and can construct the deterministic automaton with a given tree-width.
The algorithm constructs the final deterministic automaton  by running in parallel the NBA for all possible tree-widths, rather than working on SCCs independently as we do in this work.

\begin{figure}[t]
    \centering
    \includegraphics{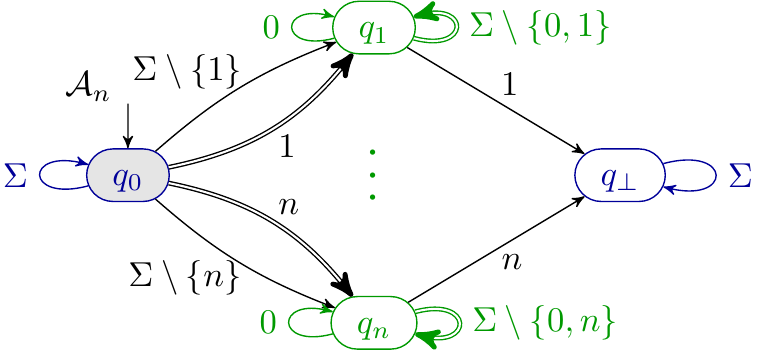}
    \caption{A family of NBAs $\aut_{n}$ with $\alphabet = \setnocond{0,1, \cdots, n}$.}
    \label{fig:ldba-family}
\end{figure}

Compared to the algorithms operating on the whole NBA, our algorithm can be exponentially better on the family of NBAs shown in Figure~\ref{fig:ldba-family}, as formalized in Theorem~\ref{thm:better-complexity-nbas};
we can encounter some variation of this family of NBAs when working with fairness properties.
The intuition is that we take care of the DACs $\setnocond{q_{i}}_{i = 1}^{n}$ independently, so for each of them we have only two choices:
either the run is in the DAC, or it is not in the DAC;
resulting in a single exponential number of combinations.
Existing works~\cite{DBLP:conf/tacas/EsparzaKRS17,DBLP:journals/fuin/Redziejowski12,safra1988complexity,DBLP:journals/lmcs/Piterman07,kahler2008complementation,DBLP:conf/icalp/LodingP19} order the runs entering the DACs based on when they visit accepting transitions, in which every order corresponds to a permutation of $\setnocond{q_{1}, \cdots, q_{n}}$.
See Appendix~\ref{app:proof-better-complexity-on-nbas} for a detailed proof.

\begin{restatable}[]{theorem}{familyOfNbasBetterComplexity}
\label{thm:better-complexity-nbas}
    There exists a family of NBAs $\aut_{n}$ with $n + 2$ states for which the algorithms in~\cite{DBLP:conf/tacas/EsparzaKRS17,DBLP:journals/fuin/Redziejowski12,safra1988complexity,DBLP:journals/lmcs/Piterman07,kahler2008complementation,DBLP:conf/icalp/LodingP19} give a DPA with at least $n!$ macrostates while ours gives a DELA with at most $2^{n + 2}$ macrostates.
\end{restatable}
In practice, for each NBA $\aut_{n}$, $n \geq 3$, \cola produces a DELA/DPA with $n$ macrostates, while both \spot and \owl give a DPA with $n! + 1$ macrostates.

\section{Conclusion and Future Work}
\label{sec:conclusion}

We proposed a divide-and-conquer determinization construction for NBAs that takes advantage of the structure of different types of SCCs and determinizes them independently.
In particular, our construction can be exponentially better than classical works on a family of NBAs.
Experiments showed that our algorithm outperforms the state-of-the-art implementations regarding the number of states and transitions on a large set of benchmarks.
To summarize, our divide-and-conquer determinization construction is very \emph{practical}, being a good complement to existing theoretical approaches.

Our divide-and-conquer approach for NBAs can also be applied to the complementation problems of NBAs.
By Proposition~\ref{prop:run-stable-scc}, $w$ is not accepted by $\aut$ if and only if there are no accepting runs staying in an SCC.
Thus we can construct a \emph{generalized} \buchi automaton with a conjunction of $\accinf{i}$ as the acceptance formula to accept the complement language $\infwords\setminus\lang{\aut}$ of $\aut$;
the generalized \buchi automaton in fact takes the \emph{intersection} of the complement language of each type of SCCs.
For complementing IWCs, we use the same construction as determinization except that the acceptance formula will be $\accinf{1}$.
For complementing DACs, we can borrow the idea of NCSB complementation construction~\cite{Blahoudek16} which complements SDBAs in time $4^{n}$.
For complementing NACs, we just adapt the \emph{slice-based} complementation~\cite{kahler2008complementation} of general NBAs.
We leave the details of this divide-and-conquer complementation construction for NBAs as future work.

\subsubsection*{Acknowledgements.}
We thank the anonymous reviewers for their valuable suggestions to this paper.
This work is supported in part by the National Natural Science Foundation of China (Grant No. 62102407 and 61836005), NSF grants IIS-1527668, CCF-1704883,
IIS-1830549, CNS-2016656, DoD MURI grant N00014-20-1-2787,
and an award from the Maryland Procurement Office.
\protect\newline
\protect\includegraphics[height=8pt]{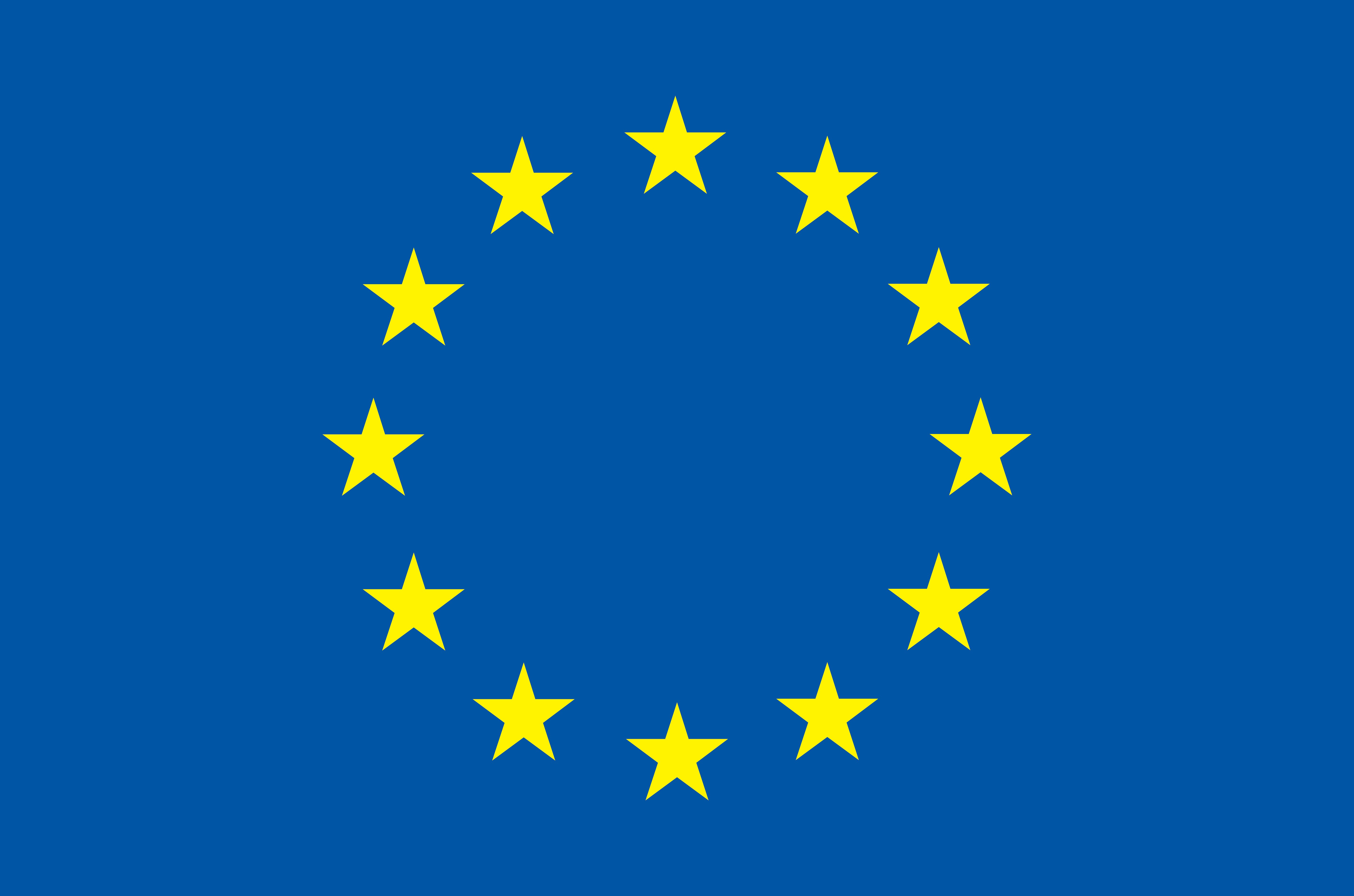} This project has received funding from the European Union’s Horizon 2020 research and innovation programme under the Marie Sk\l{}odowska-Curie grant agreement No 101008233.

\bibliographystyle{splncs04}
\bibliography{ref}

\clearpage
\newpage
\appendix

\section{Formal Proofs of the Theorems}
\label{app:proofs}

\subsection{Proof of Lemma~\ref{lem:acc-cond-iwcs}}
\label{app:proofs-acc-cond-iwcs}
\AccIWCs*

\subsubsection*{Statement~\eqref{lem:acc-cond-iwcs:correctness}: correctness.}
Assume that there is a run $\run$ that enters into an accepting IWC $C \subseteq \reachweakacc$ at level $k$ and then stays there forever.
By definition, we have that $\wordletter{\run}{\ell} \in S_{\ell}$ as well as $\wordletter{\run}{\ell} \in \reachweakacc \subseteq \reachweak$ for all $\ell \geq k$, thus also $\wordletter{\run}{\ell} \in P_{\ell}$;
recall that color $1$ is emitted solely when $O$ is empty.
If in the $(P, O)$-sequence, $O$ never becomes empty after level $k$, then we never get color $1$ anymore, so the claim follows trivially since we have got color $1$ at most $k$ times.
Suppose now that $O$ becomes empty at some level $h > k$;
this implies that $\wordletter{\run}{h+1} \in O_{h+1}$, since by definition we have $O_{h+1} = \trans(P_{h}, \wordletter{w}{h}) \cap \reachweakacc$ and $\wordletter{\run}{h+1} \in P_{h+1}$ as well as $\wordletter{\run}{h+1} \in \reachweakacc$.
Since we have $\wordletter{\run}{j} \in \reachweakacc$ for all $ j \geq h+1$ by the hypothesis that $\run$ enters into an accepting IWC $C \subseteq \reachweakacc$ at level $k$ and then stays there forever, we have that $O_{j}$
can never become empty anymore.
As before, we never get color $1$ anymore, so the claim follows since we have got color $1$ at most $h$ times.

On the other hand, if we receive color $1$ finitely many times, then by definition there are only finitely many empty $O$-sets.
There must be $k \in \naturals$ such that $O_{j} \neq \emptyset$ for all $j \geq k$.
Therefore, according to K\"onig's lemma, there must be a run $\run$ of $\aut$ over $w$ such that $\wordletter{\run}{j} \in O_{j}$ for all $j \geq k$ (the states $\wordletter{\run}{i}$ with $i < k$ are not necessarily $\reachweakacc$-states).
According to Proposition~\ref{prop:run-stable-scc}, $\run$ will end up in an SCC, which must be an accepting IWC since $O_{j} \subseteq \reachweakacc$ for all $j \geq k$.
By definition, it follows that $\run$ takes accepting transitions infinitely often, thus $\run$ is accepting.

\subsubsection{Statement~\eqref{lem:acc-cond-iwcs:complexity}: complexity.}
Regarding the number of pairs $(P, O)$, note that $O \subseteq P \subseteq \reachweak$.
Thus for each $q \in \reachweak$, either $q \in O \subseteq P$, or $q \in P$ but $q \notin O$, or $q \notin P \supseteq O$.
Therefore, the number of possible $(P, O)$ pairs is at most $3^{\size{\reachweak}}$.

\subsection{Proof of Lemma~\ref{lem:acc-run-dacs}}
\label{app:LDBAacceptingBranchStableLevelCodeterministicDAG}

\DACAcceptingRunSize*

\subsubsection*{Statement~\eqref{lem:acc-run-dacs:correctness}: correctness.}
Assume that there is an accepting run $\run$ of $\aut$ over $w$ that enters and never leaves the DAC $\reachdet$:
this means that there is $\ell \in \naturals$ such that $\wordletter{\run}{j} \in S_{j} \cap \reachdet$ for all $j \geq \ell$.
By definition, for all such $j \geq \ell$ the sequence of values $\labelling_{j}(\wordletter{\run}{j})$ never becomes $\labellingUndef$ and never increases.
This implies that such sequence eventually stabilizes, thus from some point $h > \ell$, we have $\labelling_{j}(\wordletter{\run}{j}) = k$ for all $j \geq h$;
moreover, we also have $\setnocond{1, \cdots, k} \subseteq \values{\labelling_{j}}$.
This implies that $\min \parityTranRejValues(\labelling_{j}, \wordletter{w}{j}, \labelling_{j+1}) = p > k$ after level $h$ while $\parityTranAccValues(\labelling_{j}, \wordletter{w}{j}, \labelling_{j+1})$ can be empty (resulting in the color $c = \min \setnocond{2 \cdot (\size{\reachdet} + 1), 2p - 1} > 2k$) or non-empty (resulting in the color $c = 2 \cdot \min \parityTranAccValues(\labelling_{j}, \wordletter{w}{j}, \labelling_{j+1}) \leq 2 k < 2 \cdot \size{\reachdet}$), depending on whether accepting transitions are taken.
Since the run $\run$ is accepting, we know that accepting transitions are taken infinitely often, so the minimum color occurring infinitely often is some even number at most $2k$.

Assume that the minimum color occurring infinitely often is even;
let us call it $2k$ with $k \geq 1$.
Let $h > 0$ be the level after which the minimum color we receive infinitely often is $2k$.
It is easy to see that after level $h$, the run $\run$ with the labelling number $k$ stays in $\reachdet$ and is accepting by definition.
Since $\wordletter{\run}{h}$ is a state reachable from the initial state $\init$, we then have an accepting run of $\aut$ over $w$ that stays in $\reachdet$.

\subsubsection*{Statement~\eqref{lem:acc-run-dacs:complexity}: complexity.}
Let $n_{d} = \size{\reachdet}$.
Every possible labelling function can be seen as first choosing $0 \leq k \leq n_{d}$ states and getting a permutation from those states.
So the number of possible labeling functions is
\begin{align*}
    \sum_{k = 0}^{n_{d}} \binom{n_{d}}{k} \cdot k!
    & = \sum_{k = 0}^{n_{d}} \frac{n_{d}!}{(n_{d} - k)! \cdot k!} \cdot k! \\
    & = \sum_{k = 0}^{n_{d}} \frac{n_{d}!}{(n_{d} - k)!} \\
    & \leq n_{d}! \cdot \Bigg(\frac{1}{0!} + \frac{1}{1} + \frac{1}{1 \times 2} + \frac{1}{2 \times 3} + \cdots + \frac{1}{(n_{d} - 1) \times (n_{d})}\Bigg) \\
    & = n_{d}! \cdot \Bigg(1 + 1 + \Big(\frac{1}{1} - \frac{1}{2}\Big) + \Big(\frac{1}{2} - \frac{1}{3}\Big) + \cdots + \Big(\frac{1}{n_{d} - 1} - \frac{1}{n_{d}}\Big)\Bigg) \\
    & = n_{d}! \cdot \Bigg(1 + 1 + \frac{1}{1} - \frac{1}{2} + \frac{1}{2} - \frac{1}{3} + \cdots + \frac{1}{n_{d} - 1} - \frac{1}{n_{d}} \Bigg) \\
    & = n_{d}! \cdot \Bigg(3 - \frac{1}{n_{d}}\Bigg) \\
    & \leq 3 \cdot n_{d}!.
\end{align*}
Thus the number of possible labelling functions $\labelling$ is at most $3 \cdot \size{\reachdet}!$.

\subsection{Proof of Lemma~\ref{lem:nac-acc-cond}}
\nacAccCond*

\subsubsection*{Statement~\eqref{lem:nac-acc-cond:correctness}: correctness.}
Assume that there exists an accepting run $\run $ of $\aut$ over $w$ eventually staying in $\reachnondet$.
Assume that at level $\ell_{0}$, $\run$ enters $\reachnondet$ and is assigned with labelling $[p]$.
We then have the sequence of labellings of $\run$ inside $\reachnondet$ as $\labellingtree_{\ell_{0}}(\wordletter{\run}{\ell_{0}}) \labellingtree_{\ell_{0} + 1}(\wordletter{\run}{\ell_{0} + 1}) \cdots$.

Every time $\run$ visits an accepting transition, the labelling list associated with the $\run$-state will be appended with an additional integer larger than the ones already in the labelling list.
By construction, the first integer in $\labellingtree_{\ell}(\wordletter{\run}{\ell})$ with $\ell \geq \ell_{0}$ cannot increase while reading $w$, i.e., when $\ell$ increases.
Since $\run$ stays in $\reachnondet$ forever, the first integer in each labelling list of $\run$ must stabilize at some level.
The second integer in each labelling list of $\run$ may also become stable at some point, depending on whether it will be removed due to the occurrence of good events.
In general, there must exist a maximum integer $k$ and a level $\ell_{1} > \ell_{0}$ such that $k$ occurs in $\labellingtree''_{\ell}(\wordletter{\run}{\ell})$ and $\labellingtree_{\ell}(\wordletter{\run}{\ell})$ for all $\ell > \ell_{1}$ and all integers $k' \leq k$ after level $\ell_{1}$ will not disappear;
if one of such $k'$ would disappear, $k$ will be mapped to a smaller integer via the function $\orderPosition$.

Thus, every time we are at a level $\ell > \ell_{1}$ where the maximum number in $\labellingtree_{\ell}(\wordletter{\run}{\ell})$ is larger than $k$ but the maximum integer in $\labellingtree_{\ell+1}(\wordletter{\run}{\ell+1})$ is $k$, we also need to have seen a good event;
thus we will put the integer $k$ into $\parityTranAccValues(e)$, where $e$ is the transition $e = (\labellingtree_{\ell}(\wordletter{\run}{\ell}), \wordletter{w}{\ell}, \labellingtree_{\ell+1}(\wordletter{\run}{\ell+1}))$.

Let $k'$ be the maximum number occurring in the list $\labellingtree_{\ell}(\wordletter{\run}{\ell})$;
we have states associated with $k'$ since $\run$ always continues.
Thus, by definition, we must have seen a good event represented by the integer $k$.
We then will receive an even color at most $2k$.
Moreover, it is not possible to put an integer less than $k$ in $\parityTranRejValues(e)$, since there are states associated with them, otherwise $k$ would be renumbered against the assumption that $k$ is stable.
Thus we have proved that the minimum color we receive infinitely often is even and it must be at most $2k$.

As for the other direction, we assume that the minimum color we receive infinitely often is even, say $2k$.
Assume that the sequence of labelling functions over $w$ is $\labellingtree_{0} \cdots \labellingtree_{\ell} \cdots$.
There are infinitely many transitions between two successive labelling functions where we receive the color $2k$.
Thus, by construction, this happens because there must be infinitely many labelling function $\labellingtree_{i}$ such that $(\labellingtree_{i-1}, \wordletter{w}{i-1}, \labellingtree_{i})$ is associated with the color $2k$.
Since the number of possible labelling functions is finite, there must be a labelling function, say $\labellingtree$, such that $(\labellingtree_{i-1}, \wordletter{w}{i-1}, \labellingtree = \labellingtree_{i})$ is assigned the color $2k$.
That is, the labelling function $\labellingtree$ has been visited for infinitely many times.
It follows that all the states in $\reachnondet$ associated with the color $k$ in $\labellingtree$ have been reached for infinitely many times and all runs branched from those states have all visited accepting transitions, since by definition we set all branched runs back to the labelling list whose last integer is $k$.
According to the K\"onig's lemma, there must be a run that visits accepting transitions infinitely often starting from the states in $\reachnondet$.
Since every state in $\reachnondet$ is reachable from the initial state $\init$, there must be an accepting run staying in the NAC $\reachnondet$.

\subsubsection*{Statement~\eqref{lem:nac-acc-cond:complexity}: complexity.}
For a given NAC $\reachnondet$, we can map a labelling function $\labellingtree_{\ell}$ at level $\ell$ to a Safra-tree like structure in which each node is labelled with a set of $\aut$-states.
The root node is constantly labelled with the set $\reachnondet$.
The tree has at most $\size{\reachnondet}$ nodes with the root node being $0$.
For a state $q \in S_{\ell}\cap \reachnondet$, $\labellingtree_{\ell}(q)$ is in fact a list of node numbers where the state $q$ resides in.
We say that the maximal number of the labelling $\labellingtree_{\ell}(q)$ is the host node of $q$.
We denote by $\snode(q)$ the host node of $q$.
By definition, the integers in a labelling are already in ascending order.
Therefore, we have the following properties:
\begin{itemize}
\item
    if $\labellingtree_{\ell}(q)$ is a prefix of $\labellingtree_{\ell}(q')$, then $\snode(q)$ is an ancestor node of $\snode(q')$.
    Moreover, we have $\snode(q) < \snode(q')$.
    In particular, if $\labellingtree_{\ell}(q)$ has one less integer than $\labellingtree_{\ell}(q')$, then $\snode(q)$ is the parent of $\snode(q')$;
\item
    the states labelled in a parent node $n$ must form a superset of all states in its children, just like that in history trees~\cite{Schewe09};
    and
\item
    the states in two different child nodes of a node are disjoint.
    This is immediate by construction.
\end{itemize}
Also, the node number of a parent node must be smaller than that of its every child~\cite{Schewe09,DBLP:journals/lmcs/Piterman07} according to the order of entering $\reachnondet$.
(This is called later introduction record in literature.)
The labelling of the tree with the states can be seen as a function from $\reachnondet$ to $\setnocond{1, \cdots, m}$, where $m$ is the number of nodes, that maps a state $q \in \reachnondet$ to the largest node number it resides in and to $0$ if the state is not reached in the current level, thus it is in the root node.
According to~\cite{Schewe09} (in the second paragraph of page 179), the number of different trees is $2 \cdot (\size{\reachnondet} - 1)! \cdot \size{\reachnondet}! \leq 2 \cdot (\size{\reachnondet}!)^{2}$.

\subsection{Proof of Theorem~\ref{thm:LDBAlanguageSizeParityAutomaton}}
\label{app:LDBAlanguageSizeParityAutomaton}

\LDBAlanguageSizeParityAutomaton*

\subsubsection*{Statement~\eqref{thm:LDBAlanguageSizeParityAutomaton:correctness}: correctness.}
Each word $w$ accepted by $\aut$, by Proposition~\ref{prop:run-stable-scc}, eventually gets trapped in an accepting SCC, i.e., either by the IWCs, or a DAC or a NAC;
thus the corresponding results given in  Lemmas~\ref{lem:acc-cond-iwcs},~\ref{lem:acc-run-dacs}, and~\ref{lem:nac-acc-cond} apply, respectively.
Since $\dela$ accepts the union of the words accepted by each single construction, we have that $w$ is also accepted by $\dela$, that is, $\lang{\aut} \subseteq \lang{\dela}$.

On the other hand, if a word $w$ is accepted by $\dela$, then it must be accepted by one of its components, i.e., either the one for IWCs, for DACs or for NACs;
thus the corresponding results given in  Lemmas~\ref{lem:acc-cond-iwcs},~\ref{lem:acc-run-dacs}, and~\ref{lem:nac-acc-cond} apply, respectively.
In all cases, we derive that $w$ is accepted by $\aut$, that is, $\lang{\dela} \subseteq \lang{\aut}$.
By combining the obtained language inclusions, we derive $\lang{\dela} = \lang{\aut}$ as required.

\subsubsection*{Statement~\eqref{thm:LDBAlanguageSizeParityAutomaton:complexity}: complexity}
The size of $\dela$ follows directly by taking the product of the number of pairs or labelling functions given by Lemmas~\ref{lem:acc-cond-iwcs},~\ref{lem:acc-run-dacs}, and~\ref{lem:nac-acc-cond};
the number of colors is just by definition.

\subsection{Proof of Corollary~\ref{coro:worst-case-subclasses-determinization}}
\label{app:results-subclasses-nbas}

\worstCaseDeterminizationSubclass*

\subsubsection*{Statement~\eqref{coro:worst-case-subclasses-determinization:weak}: weak BAs.}
The $3^{n}$ macrostates for the weak BAs case is a trivial consequence of Theorem~\ref{thm:LDBAlanguageSizeParityAutomaton}:
since a weak BA has no DACs or NACs, only IWCs, we have $\reachweak = \states$ and $d = k = 0$ that give $3^{n}$ when instantiating the formula for the number of macrostates in Theorem~\ref{thm:LDBAlanguageSizeParityAutomaton}.

\subsubsection*{Statement~\eqref{coro:worst-case-subclasses-determinization:elevator}: elevator BAs.}
We now prove that determinizing elevator BAs can be done in $\bigM(n!)$;
it is in fact a direct result of Theorem~\ref{thm:LDBAlanguageSizeParityAutomaton}.
It is trivial to see that the constructed DELA has the same language as the input, according to Theorem~\ref{thm:LDBAlanguageSizeParityAutomaton}, so we just need to prove that the determinization complexity is in $\bigM(n!)$.

Recall that elevator BAs have only IWCs and DACs as SCCs, thus $k = 0$.
Let $n_{w} = \size{\reachweak}$ and $n_{i} = \size{\reachdet^{i}}$ for each of the $d$ DACs $\reachdet^{i}$.
In this case, the worst-case complexity of our algorithm is only
\begin{align*}
    3^{n_{w}} \cdot \prod_{i = 1}^{d} 3 \cdot n_{i}!
    & = 3^{n_{w} + d} \cdot (n_{1} + \cdots + n_{d})! \cdot \frac{1}{\binom{n_{1} + \cdots + n_{d}}{n_{1}}} \cdot \frac{1}{\binom{n_{2} + \cdots + n_{d}}{n_{2}}} \cdot \cdots \cdot \frac{1}{\binom{n_{d}}{n_{d}}} \\
    & \leq 3^{n}  \cdot n! \cdot \prod_{i=1}^{d} \frac{1}{n_{i}} \in \bigO(n!),
\end{align*}
where $n = n_{w} + \sum_{i = 1}^{d} n_{i}$.
It is already known~\cite{DBLP:conf/fsttcs/Loding99,DBLP:conf/tacas/EsparzaKRS17} that the lower bound for determinizing semi-deterministic \buchi automata (SDBAs), a strict subset of elevator NBAs, is $n!$.
Thus our determinization algorithm for elevator \buchi automata is already asymptotically optimal since the lower bound and the upper bound coincide as $\bigM(n!)$.

\subsection{Proof of Theorem~\ref{thm:dela-to-dra}}

\DELAtoDRA*
The different algorithms presented in Section~\ref{sec:determinizationSCCs}, that form the basis for the construction of $\dela$ in Section~\ref{sec:LDBADeterminizationToParity}, generate a parity condition for the acceptance inside the various SCCs.

It is known that a DPA $\dpa$ can be trivially transformed into a Rabin automaton $\dra$ by changing only the accepting formula $\acc$ and the coloring function $\parity$, that is, states and transitions remain the same (cf.~\cite[Transformation~1.18]{DBLP:conf/dagstuhl/Farwer01});
we adopt a similar approach for our case, just we have to consider the transpose operation we applied to the colors and the fact that the global acceptance condition $\delaAcc$ is the disjunction of the parity conditions of the various SCCs.

By definition, a word $w$ is accepted by a parity automaton with color set $\colorsymbol{k}$ if there is a run $\run$ over $w$ such that $\min \parity(\inftimes{\run})$ is even, that is, $p$ is the minimum even color occurring infinitely often and all other colors $p' < p$ occur only finitely often.
To define the Rabin acceptance formula, we just give all transitions originally with color $p$ the color $p/2$, and all transitions with smaller value the color $p/2 - 1$;
then the Rabin acceptance formula is $\acc = (\accfin{0} \land \accinf{1}) \lor \cdots \lor (\accfin{k/2-1} \land \accinf{k/2})$ and the color set is $\colorsymbol{k/2}$ (note that $k$ is an even number by definition).

From the definition of the acceptance formula $\delaAcc$, we can see that it is a disjunction of parity-like acceptance conditions.
So it follows that we can also translate $\dela$ to a DRA without blow-up of states and transitions.

\subsection{Proof of Theorem~\ref{thm:better-complexity-nbas}}
\label{app:proof-better-complexity-on-nbas}

\familyOfNbasBetterComplexity*

Each NBA $\aut_{n} = (\states, q_{0}, \trans, \accset)$ in the family is like the one depicted in Figure~\ref{fig:ldba-family}.

Consider our construction:
for the NBA $\aut_{n}$, there are $n$ DACs $\reachdet^{i} = \setnocond{q_{i}}$, for $1 \leq i \leq n$.
The SCCs $\setnocond{q_{0}}$ and $\setnocond{q_{\bot}}$ are both nonaccepting IWCs.
So in our algorithm, the DELA we construct has at most $2^{2} \times 2^{n} = 2^{n+2}$ macrostates:
$2^{2}$ pairs $(P, O)$ for the IWCs, since $\reachweakacc = \emptyset$ thus the $O$-sets for the IWCs are always empty;
$2^{n}$ choices for the DAC level encoding, since for each DAC $\reachdet^{i} = \setnocond{q_{i}}$, we have $2$ choices:
either no state or only $q_{i}$ is currently visited in $\reachdet^{i}$.
In this way we can give a more precise bound than the general one provided by Theorem~\ref{thm:LDBAlanguageSizeParityAutomaton}.

Consider now the constructions from literature:
the deterministic part of $\aut_{n}$ is $\dstates = \setnocond{q_{1}, \cdots, q_{n}, q_{\bot}}$;
the algorithms described in~\cite{DBLP:conf/tacas/EsparzaKRS17,DBLP:journals/fuin/Redziejowski12,safra1988complexity,kahler2008complementation,DBLP:journals/lmcs/Piterman07,DBLP:conf/icalp/LodingP19} give precedence to the runs that see an accepting transition earlier and visit more accepting transitions.
That is, when reading the letter $i \in \setnocond{1, \cdots, n}$, the order of the runs entering in $\dstates$ is $q_{i} < q_{1} < \cdots < q_{i-1} < q_{i+1} < \cdots < q_{n}$ if $i \neq 1$ and $q_{1} < \cdots < q_{n}$ otherwise.

To prove that the DPA constructed by~\cite{DBLP:conf/tacas/EsparzaKRS17,DBLP:journals/fuin/Redziejowski12,safra1988complexity,kahler2008complementation,DBLP:journals/lmcs/Piterman07,DBLP:conf/icalp/LodingP19} has at least $n!$ macrostates, we show how to ensure that the order of the states $D = \setnocond{q_{1}, \cdots, q_{n}}$ in a reachable macrostate is a permutation of the states in $D$;
since there are $n!$ permutations, there need to be at least $n!$ macrostates.
To this end, we assume that we are given a permutation $\pi = q_{i_{1}} < q_{i_{2}} < q_{i_{3}} < \cdots < q_{i_{n}}$ and we show how to reach the corresponding macrostate from the initial state of the DPA via a finite word $u_{\pi}$, by constructing such a word $u_{\pi}$ from the above permutation $\pi$.
Note that we ignore the place where we have $q_{\bot}$ and only focus on the order of the $D$-states.

From the initial state of the DPA, which contains only the initial state $q_{0}$ of $\aut_{n}$, to put $q_{i_{1}}$ in the first place, we first read the letter $i_{1}$, which leads to the macrostate with $q_{i_{1}} < q_{1} < \cdots q_{i_{1} - 1} < q_{i_{1} + 1} < \cdots < q_{n}$.
Note that we see an accepting transition from $q_{0}$ to $q_{i_{1}}$ via $i_{1}$, so the algorithm gives precedence to the run that reaches $q_{i_{1}}$ and orders the other states according to the usual state order $\stateOrder$.
To put $q_{i_{2}}$ in the second position, we can just read once each letter from $\setnocond{1, \cdots, n} \setminus \setnocond{i_{1}, i_{2}}$ in ascending order.
This makes all states in $D$, except for $q_{i_{1}}$ and $q_{i_{2}}$, transition to $q_{\bot}$, leaving $q_{i_{2}}$ just after $q_{i_{1}}$, obtaining the order $q_{i_{1}} < q_{i_{2}} < q_{1} < \cdots < q_{n}$.
Note that states will keep coming from the initial state $q_{0}$ and if the newly arrived states are already present, they will be ignored.
To put $q_{i_{3}}$ in the third position, similarly to the previous case we read all letters from $\setnocond{1, \cdots, n} \setminus \setnocond{i_{1}, i_{2}, i_{3}}$ in ascending order.
By proceeding in the same way, we obtain the order $q_{i_{1}} < q_{i_{2}} < q_{i_{3}} < \cdots < q_{i_{n}}$, which is exactly the given permutation $\pi$.
Given the arbitrary choice of $\pi$, we have that all permutations correspond to a different macrostate in the DPA constructed by the algorithms given in~\cite{DBLP:conf/tacas/EsparzaKRS17,DBLP:journals/fuin/Redziejowski12,safra1988complexity,kahler2008complementation,DBLP:journals/lmcs/Piterman07,DBLP:conf/icalp/LodingP19}, so the DPA has at least $n!$ macrostates.

\subsubsection{Consistent experiments.}
To check whether the behavior of the actual implementations of the algorithms given in~\cite{DBLP:conf/tacas/EsparzaKRS17,DBLP:journals/fuin/Redziejowski12,safra1988complexity,kahler2008complementation,DBLP:journals/lmcs/Piterman07,DBLP:conf/icalp/LodingP19} are consistent with the bounds given by Theorem~\ref{thm:better-complexity-nbas}, we have performed some experiments on the family of NBAs with \cola, \spot and \owl, finding that they behave consistently with Theorem~\ref{thm:better-complexity-nbas}.
In practice, for the given NBA $\aut_{n}$, with $n \geq 3$, \cola gives a DPA with $n$ macrostates and $\accfin{0} \land \accinf{1}$ as acceptance formula;
\spot returns a DPA with $n! + 1$ macrostates and the same acceptance formula $\accfin{0} \land \accinf{1}$;
and
\owl produces a DPA with also $n! + 1$ macrostates while the acceptance formula is $\accinf{0} \lor (\accfin{1} \land (\accinf{2} \lor \accfin{3}))$;
however it fails on $\aut_{7}$ and larger automata with a \texttt{java.lang.StackOverflowError} when it tries to optimize the constructed automaton.
On the other hand, \cola can go up for a very large $n$, while \spot already timeouts on $\aut_{11}$.
We remark that the acceptance formulas from \cola and \spot have been simplified by the algorithms implemented in \spot.

\section{Additional Plots for the Experiments}
\label{app:plotsExperiments}

In this appendix we provide additional plots for the experiments presented in Section~\ref{sec:experiments}, which we omitted in the main part of the paper to keep it short.

\subsection{Determinization of NBAs from \automatabenchmarks}

\begin{figure}[t]
    \centering
    \resizebox{\linewidth}{!}{
    \includegraphics{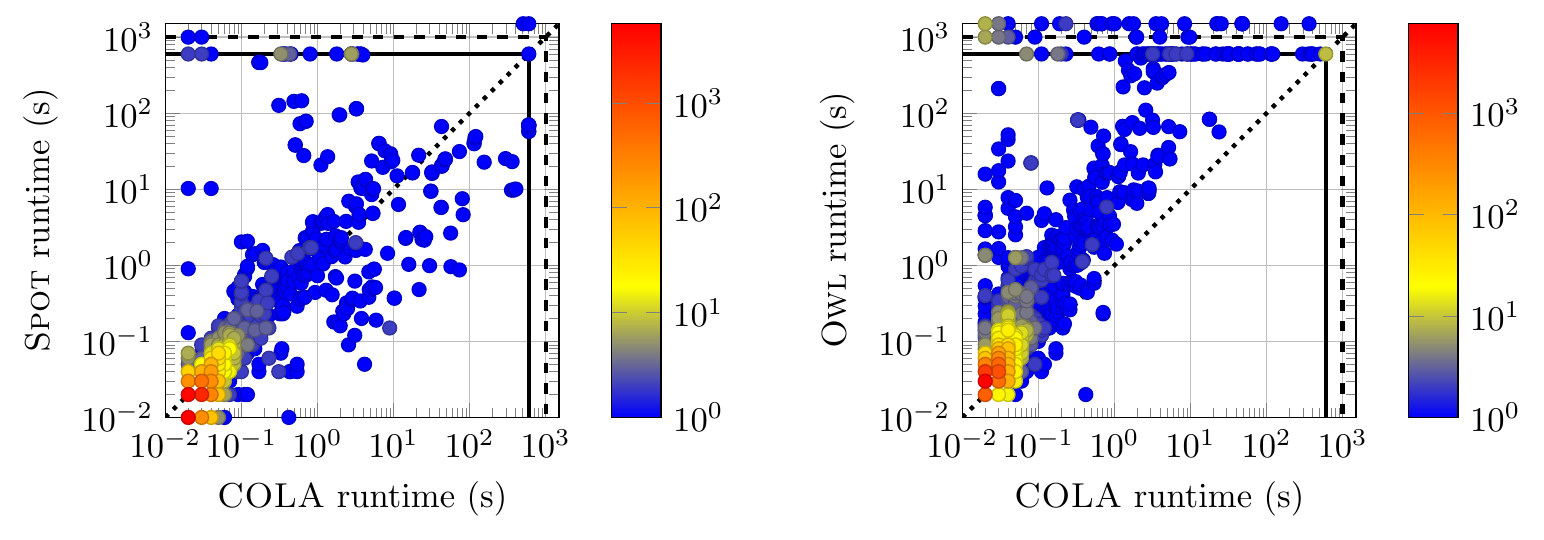}
    }
    \caption{Runtime comparison on the determinization of \automatabenchmarks.}
    \label{fig:experimentsAutomataParityPlottime}
\end{figure}
In Figure~\ref{fig:experimentsAutomataParityPlottime} we provide scatter plots showing the detailed comparison of \cola with \spot and \owl with respect to the running time on the single cases.
In the plots, the solid line just before $10^{3}$ indicates the timeout (set to $600$ seconds);
the dashed line at $10^{3}$ an out of memory result, and the border of the plot other failures.
As we can see from the plots, \owl is usually slower than \cola, except for few cases requiring less than one second.
Regarding \spot, it is frequently faster than \cola, showing how well it performs.
\cola, on the other hand, is better than \spot in several cases, on which it is faster or even able to produce a DPA within 10 seconds while \spot goes timeout.
By comparing these results with the cactus plot in Figure~\ref{fig:experimentsAutomataParityCactusPlottime}, we can conclude that there is no clear winner between \spot and \cola when we consider just the running time of the tools.

\begin{figure}[t]
    \centering
    \resizebox{\linewidth}{!}{
    \includegraphics{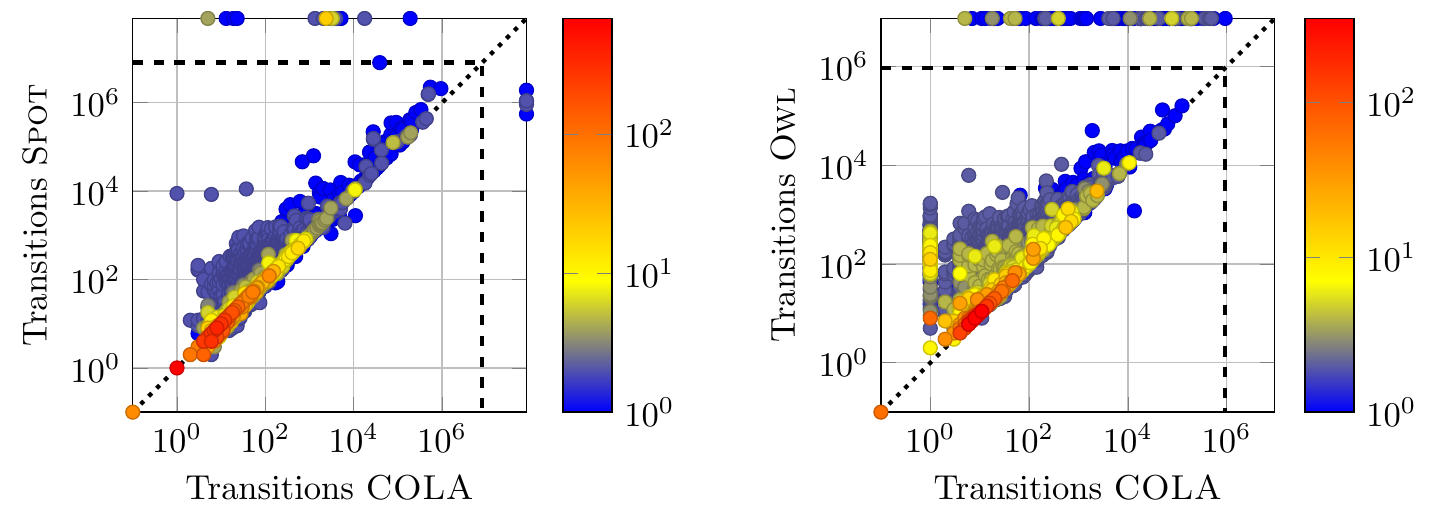}
    }
    \caption{Transitions comparison on the determinization of \automatabenchmarks; a mark at $10^{-1}$ represents automata without transitions.}
    \label{fig:experimentsAutomataParityTransitions}
\end{figure}
In Figure~\ref{fig:experimentsAutomataParityTransitions} we complete the comparison of \cola with \spot and \owl by considering the number of transitions in the DPAs generated from the NBAs available in \automatabenchmarks.
Since we use logarithmic axes and automata can have zero transitions (this happens for the automata accepting the empty language, since the HOA format allows the tools to specify automata that are not complete), in the plots we place a mark at $10^{-1}$ to represent the fact that the corresponding tool has returned an automaton with no transitions at all.
As already mentioned in the main part of the paper, the plots are very similar to the ones about the states shown in Figure~\ref{fig:experimentsAutomataParityStates}.
Regarding the average number of transitions generated by the tools, on the 15,710 commonly solved cases, we have 167 transitions on average by \cola, 266 by \spot, and 292 by \owl.
By splitting the comparison, on the 15,854 cases solved by both \cola and \spot, we have on average 900 transitions from \cola and 2,174 from \spot;
on the 15,749 common cases for \cola and \owl, we have 173 transitions from \cola and 302 from \owl.

\subsection{Determinization of NBAs from LTL Formulas}

We now present the plots about the detailed comparison of the different tools on the NBAs generated from the LTL formulas considered in~\cite{DBLP:conf/atva/KretinskyMS18}, as available in the \owl's repository.
As said in the main part of the paper, we used the \spot's \texttt{ltl2tgba} command to convert them to NBAs:
out of 645 formulas, \texttt{ltl2tgba} failed to provide the NBA for just one formula.

\begin{figure}[t]
    \centering
    \includegraphics{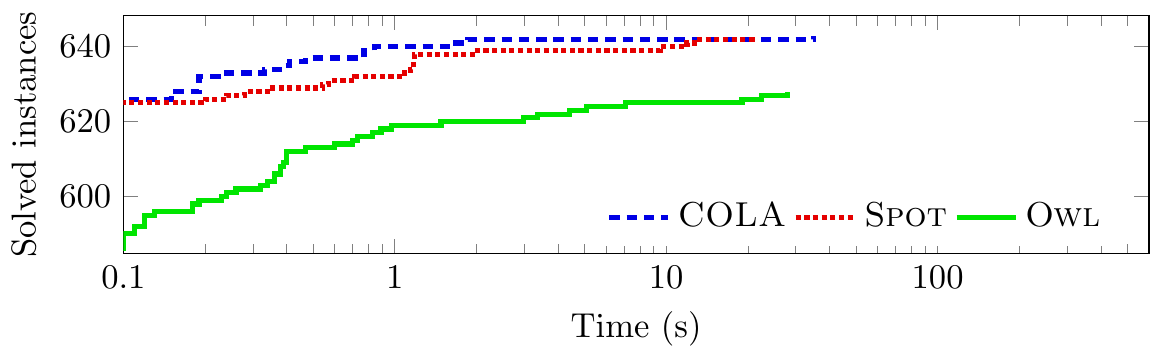}
    \caption{The cactus plot for the determinization of NBAs from LTL formulas.}
    \label{fig:experimentsFormulasNBAsParityCactus}
\end{figure}
The cactus plot shown in Figure~\ref{fig:experimentsFormulasNBAsParityCactus} is the counterpart of the one about \automatabenchmarks shown in Figure~\ref{fig:experimentsAutomataParityCactusPlottime}.
We can see that the two plots show the same trend:
\cola slightly above \spot, with \owl quite below them.
If we look at the actual numbers, we have that both \cola and \spot are able to determinize 643 cases, while \owl only 628, with a common failure for all tools.

\begin{figure}[t]
    \centering
    \resizebox{\linewidth}{!}{
    \includegraphics{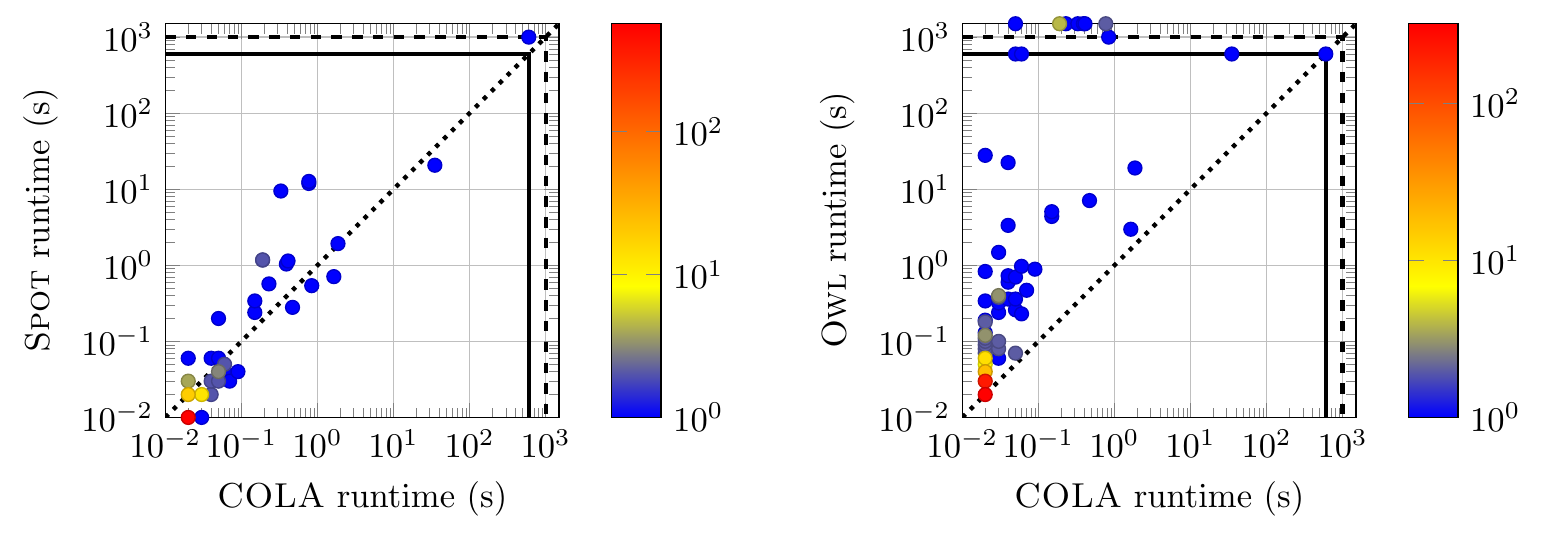}
    }
    \caption{Runtime comparison on the determinization of NBAs from LTL formulas.}
    \label{fig:experimentsFormulasNBAsParityPlottime}
\end{figure}

The plots in Figure~\ref{fig:experimentsFormulasNBAsParityPlottime} are relative to the running time of the different tools.
By comparing them with the corresponding plots in Figure~\ref{fig:experimentsAutomataParityPlottime}, we can see that they are rather similar.
For the NBAs in this set of benchmarks, however, we can note that \cola is usually faster than \spot, in particular for the inputs requiring more than $0.1$ seconds to be determinized, as already hinted by the cactus plot shown in Figure~\ref{fig:experimentsFormulasNBAsParityCactus}.

\begin{figure}[t]
    \centering
    \resizebox{\linewidth}{!}{
    \includegraphics{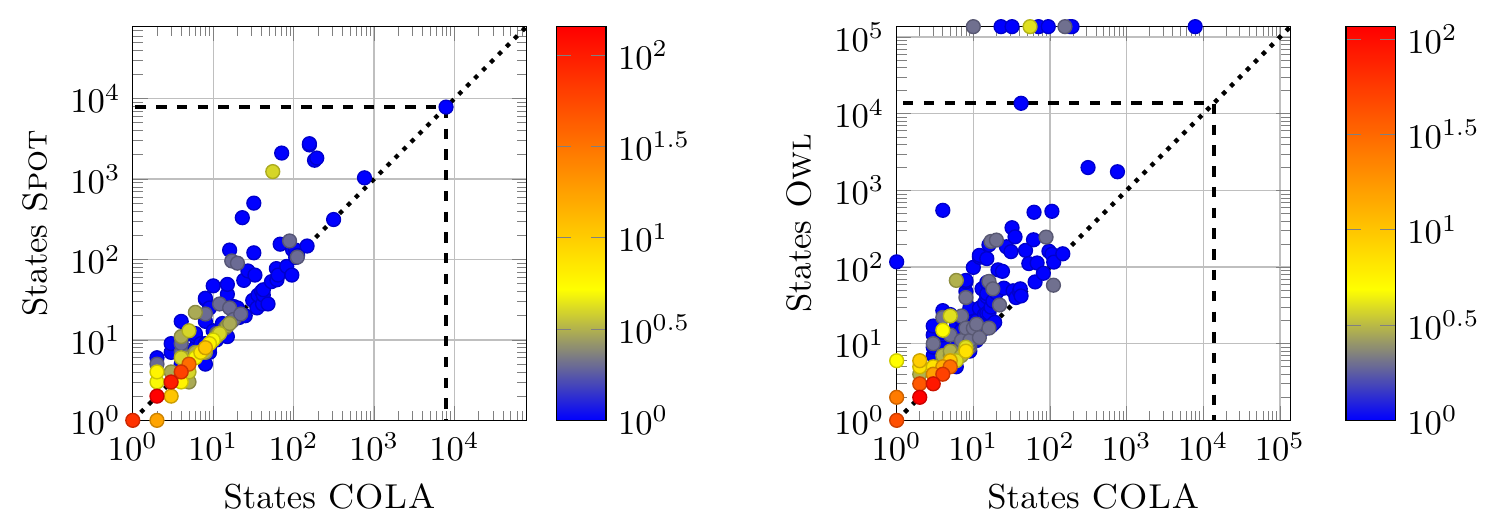}
    }
    \caption{States comparison on the determinization of NBAs from LTL formulas.}
    \label{fig:experimentsFormulasNBAsParityStates}
\end{figure}

\begin{figure}[t]
    \centering
    \resizebox{\linewidth}{!}{
    \includegraphics{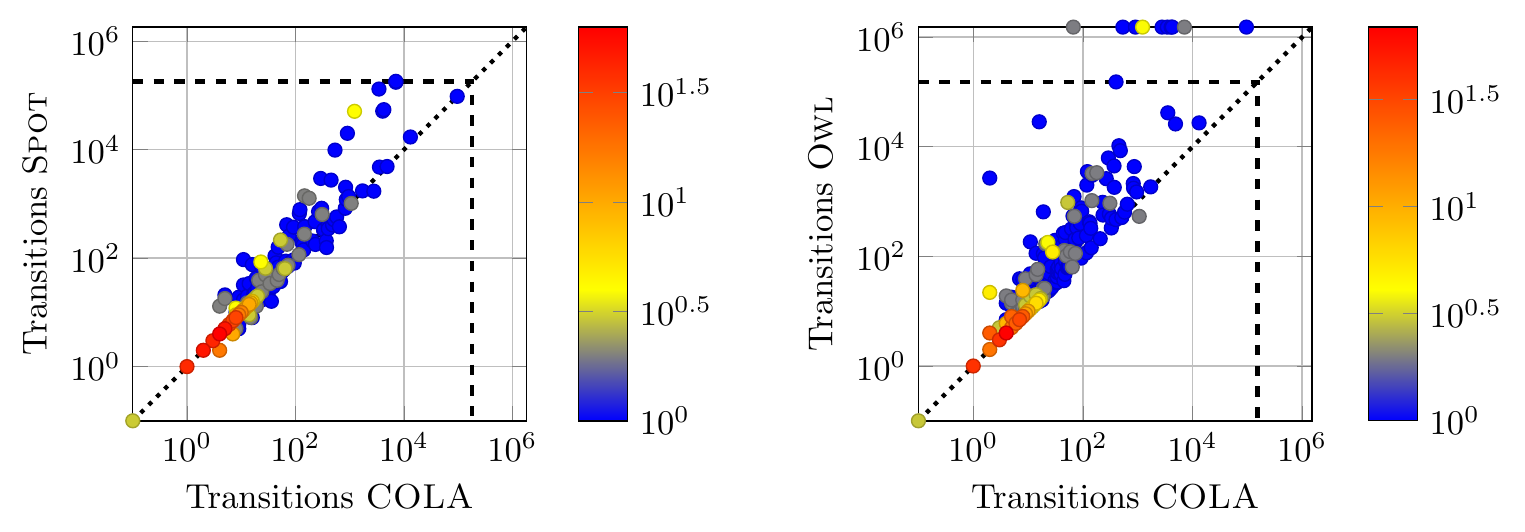}
    }
    \caption{Transitions comparison on the determinization of NBAs from LTL formulas.}
    \label{fig:experimentsFormulasNBAsParityTransitions}
\end{figure}

The scatter plots in Figure~\ref{fig:experimentsFormulasNBAsParityStates} and Figure~\ref{fig:experimentsFormulasNBAsParityTransitions} are relative to the number of states and transitions, respectively, of the generated DPAs.
Similarly to the plots about the running time, we again have that \cola and \spot produce automata of similar size, with \cola more frequently generating smaller automata than \spot in this benchmark than in \automatabenchmarks (cf. Figure~\ref{fig:experimentsAutomataParityStates} and Figure~\ref{fig:experimentsAutomataParityTransitions}, respectively).
By looking at the actual number of states, we have that on the 628 commonly solved cases, on average the DPA generated by \cola has 9 states, by \spot 11, and by \owl 45.
If we just compare \cola and \spot on all 643 solved cases, we have that \cola generates 22 states while \spot 49, on average.
This situation is similar also for the average number of transitions:
for the 628 cases solved by all tools, we have 72 for \cola, 108 for \spot, and 609 for \owl;
for the 643 cases solved by \cola and \spot, we have 275 and 1547 transitions on average, respectively.

\begin{figure}[t]
    \centering
    \resizebox{\linewidth}{!}{
    \includegraphics{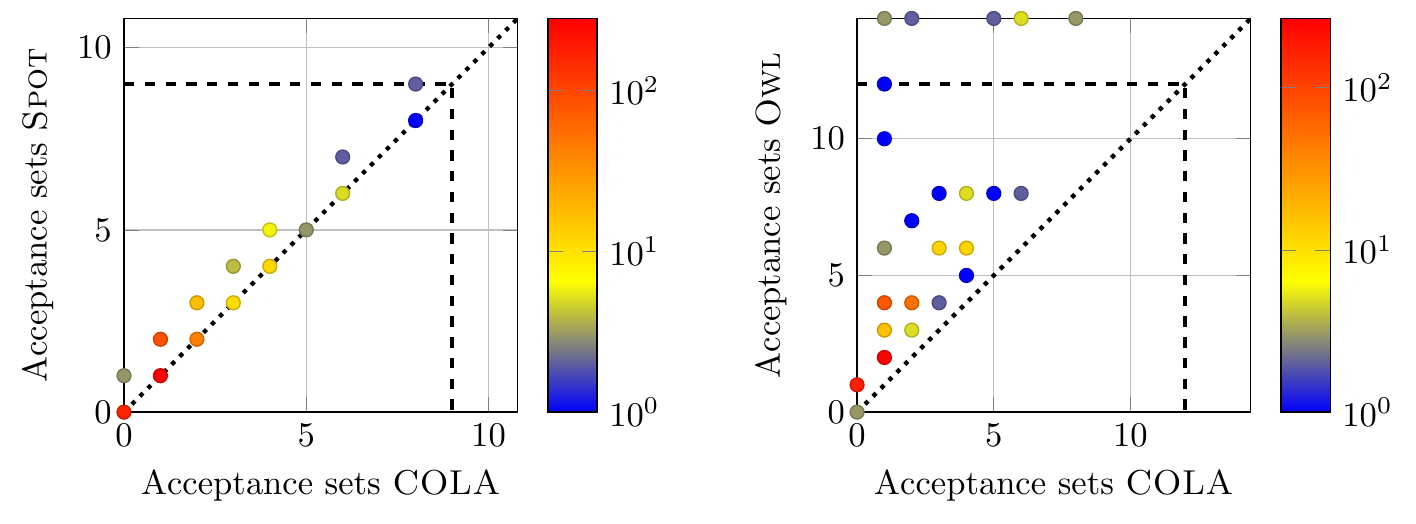}
    }
    \caption{Acceptance sets comparison on the determinization of NBAs from LTL formulas.}
    \label{fig:experimentsFormulasNBAsParityAcceptanceSets}
\end{figure}

Lastly, the plots in Figure~\ref{fig:experimentsFormulasNBAsParityAcceptanceSets} are relative to the number of colors in the generated DPAs.
Differently from the \automatabenchmarks plots shown in Figure~\ref{fig:experimentsAutomataParityAcceptanceSets}, on the NBAs from LTL formulas \cola always produces a DPA with at most the number of acceptance sets/colors of the corresponding one produced by \spot;
more precisely, \cola generates 118 times fewer colors than \spot and 525 time the same number of colors.
\owl, on the other hand, behaves better than on \automatabenchmarks, but it still generates more acceptance sets/colors than \cola, with just 3 cases (out of 628) having the same number of colors for both tools.

\end{document}